\documentclass[namedreferences,hyperref,optionalrh,solaromanenum]{spr-sola}
\usepackage{lmodern} 
\usepackage{graphicx}                    
\usepackage{color}                       
\usepackage{amsmath}
\usepackage{comment}

\chardef\us=`\_

\begin{document}

\begin{frontmatter}

\title{{Flare energetics, CME launch and heliospheric propagation for the May 2024 events, as derived from ensemble MHD modelling}  
}

%
 \author[addressref={aff1,aff2,aff3},corref,email={brigitte.schmieder@obspm.fr}]{\inits{B.}\fnm{Brigitte}~\snm{Schmieder}}
 \author[addressref={aff4,aff2},corref,email={jinhan.guo@nju.edu.cn}]{\inits{J.}\fnm{Jinhan}~\snm{Guo}}
 \author[addressref={aff5,aff6},corref,email={guillaume.aulanier@lpp.polytechnique.fr}]{\inits{G.}\fnm{Guillaume}~\snm{Aulanier}}
  \author[addressref={aff2},corref,email={anwesha.maharana@kuleuven.be}]{\inits{A.}\fnm{Anwesha}~\snm{Maharana}}
  \author[addressref={aff2,aff7},corref,email={stefaan.poedts@kuleuven.be}]{\inits{S.}\fnm{Stefaan}~\snm{Poedts}}
%
\address[id=aff1]{Observatoire de Paris, LIRA, UMR8254 (CNRS), F-92195 Meudon Principal Cedex, France} 
\address[id=aff2]{Centre for mathematical Plasma Astrophysics, Dept. of Mathematics, KU Leuven, 3001 Leuven, Belgium} 
\address[id=aff3]{LUNEX EMMESI Institut, SBIC, Kapteyn straat 1, Noordwijk 2201 BB, Netherlands}
\address[id=aff4]{School of Astronomy and Space Science and Key Laboratory of Modern Astronomy and Astrophysics, Nanjing University, Nanjing 210023, PR China}
\address[id=aff5]{Sorbonne Universit\'e, Observatoire de Paris - PSL, École Polytechnique, Institut Polytechnique de Paris,  CNRS, Laboratoire de physique des plasmas (LPP), 4 place Jussieu, F-75005 Paris, France }
\address[id=aff6]{Rosseland Centre for Solar Physics, University of Oslo, P.O. Box 1029 Blindern, N-0315 Oslo, Norway}
\address[id=aff7]{Institute of Physics, University of Maria Curie-Skłodowska, ul. Radziszewskiego 10, 20-031 Lublin, Poland}

\begin{abstract}
Many questions must be answered before understanding the relationship between the emerging magnetic flux through the solar surface and the extreme geoeffective events.  
{The main ingredients for getting X-ray class flares and large interplanetary Coronal Mass Ejections { (CMEs)} are the build-up of  electric current in the corona, }
the existence of magnetic free energy, magnetic energy/helicity ratio, twist, and magnetic stress in active regions (ARs). The upper limit of solar energy in the space research era, as well as the potential for experiencing superflares and extreme solar events, can be predicted using MHD simulations of CMEs.
\\
{ To address this problem, we consider the recent events of May 2024  and use 
 three  MHD  models:\\
1) OHM ("Observationally driven High order scheme Magnetohydrodynamic code") for investigating the magnetic evolutions at a synthetic dipole structure.\\
2) TMF (time-dependent magneto-friction) for setting up an initial non-potential magnetic field in the
active region. A zero-beta MHD model for tracing the magnetic evolution of active regions.\\
3) EUHFORIA (''European heliospheric forecasting information asset'') for interplanetary CME propagations.

For the eruptive flares with CMEs 
magnetic solar energy is computed 
along with data-constrained MHD simulations for the May 2024 events. We show the consistency between the data-initiated realistic simulation of the May 2024 big event and energy scalings from an idealised simulation of a bipolar eruption using OHM.}
The estimated free magnetic energy did not surpass { $5.2 \times 10^{32}\;$erg.} Good arrival time predictions ($<3$ hours)  are achieved with the EUHFORIA simulation with the cone model.  We note the interest in coupling all the chains of codes from the Sun to the Earth and developing different approaches to test the results.
\end{abstract}

%

\end{frontmatter}

%
 \section{Introduction}
Extreme solar events are characterised by a significant amount of magnetic energy release in the heliosphere. It can occur through electromagnetic radiation, such as X-class flares, or by kinetic energy, as with coronal mass ejections (CMEs), or by the ejection of accelerated particles. 
 \citet{Gosling1994} argued that solar flares play no fundamental
role in causing geomagnetic disturbances and expressed that there is no fundamental association between flares and CMEs. He sparked a considerable and controversial debate.
The weak relationship between flares and CMEs for medium events was confirmed later \citep{Andrews2003,Schmieder2020}.  However,
most big events have intense magnetic flares, fast CMEs, and very energetic accelerated particles, { called the Big Flare Syndrome by \citet{Kahler1982}}. The flare positions with respect to the CME spans are found in 
the centre of the CME span. The dimmings are at the edge of the flare positions \citep{Yashiro2005}.
The Carrington event in 1859 is the strongest event so far that affected telecommunications \citep{Cliver2022}. Such an event now { would} damage more technical instruments and spacecraft due to the high technology in our era. Solar flares result from the abrupt release of free magnetic energy that has previously been stored in the coronal magnetic field by flux emergence and surface motions \citep{Forbes2006}. Most of the strongest flares are eruptive, accompanied by CMEs \citep{Schrijver2009,Emslie2004}. The standard model CSHKP {\citep{Carmichael1964,Sturrock1966,Hirayama1974,Kopp1976}} attributes the flare energy release to magnetic reconnection that occurs during the escape of CMEs  \citep{Shibata1995,Lin2000,Priest2002,Moore2001}.
\citet{Schrijver2012} concluded that because the likelihood of flares larger than approximately X~3.0 remains empirically unconstrained, and based on records of sunspots and solar flares from the past four centuries, there is a 10 per cent chance of a flare being larger than about X30 in the next 30 years. Hence, the opportunity to observe a super solar flare from the Earth as energetic as in other stars is small \citep{2012Natur.485..478M}.

\begin{figure} 
\centerline{\includegraphics[width=1.\textwidth,clip=]{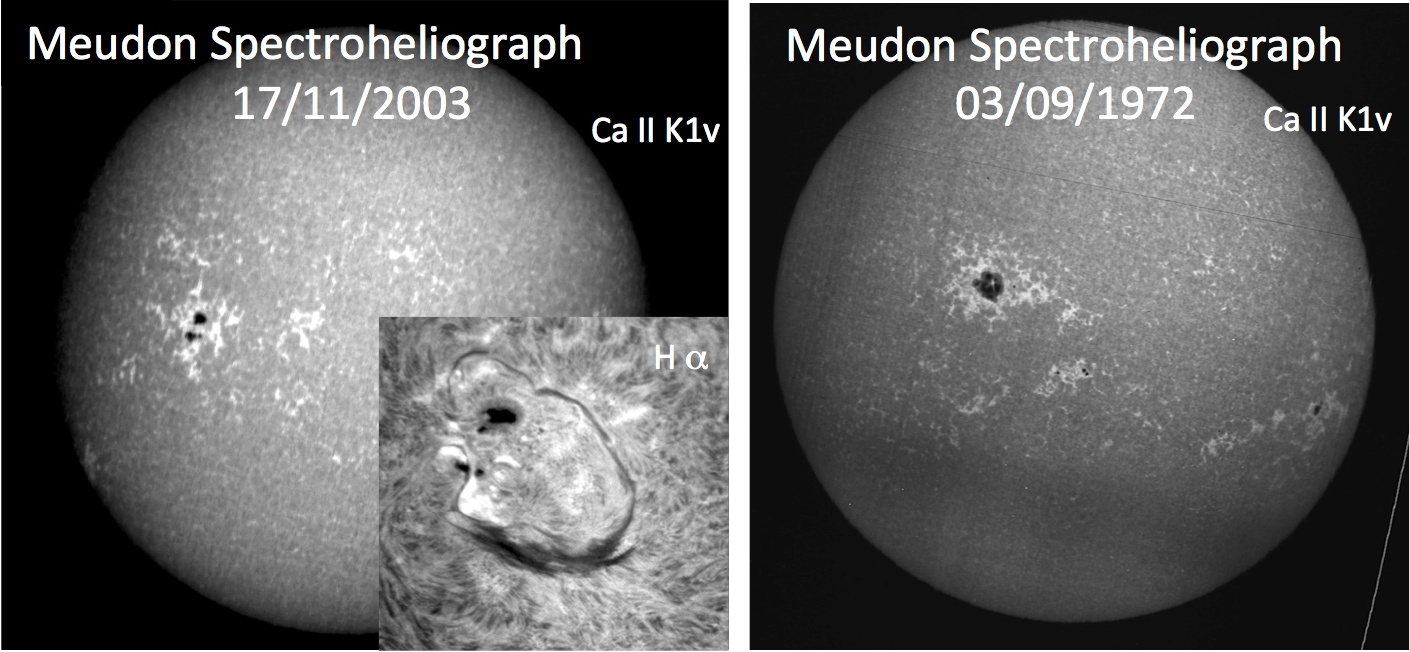}}
\caption{Spectroheliograms in Ca II  showing large sunspots obtained at the Observatoire de Paris, Meudon, leading to extreme events in 2003  with an inserted  H$\alpha$ image showing the erupted filament from Nanintal (courtesy of Ramesh Chandra) (left panel) and in  1972 (right panel).}
\label{1972}
\end{figure}

The most energetic event in the solar cycle {23} was the 20 November 2003 event, after the series of Halloween events, for which the Disturbance Storm {time } (Dst) index was not as strong as that of the November event (Dst= - 422~nT) \citep{Cliver2022}. The sunspot was one of the largest of the solar cycle but smaller than the historical event of 1972  (Figure \ref{1972}) \citep{Schmieder2018}. For the November 2003 event, it was revealed that its geoeffectiveness was mainly due to the eruption of a filament, rather than the flare \citep{Chandra2010}. Filament eruptions can have a significant impact, producing geoeffective coronal mass ejections (CMEs) and accelerating particles \citep{Schmieder2015}. 

\begin{figure} 
\centerline{\includegraphics[width=1\textwidth,clip=]{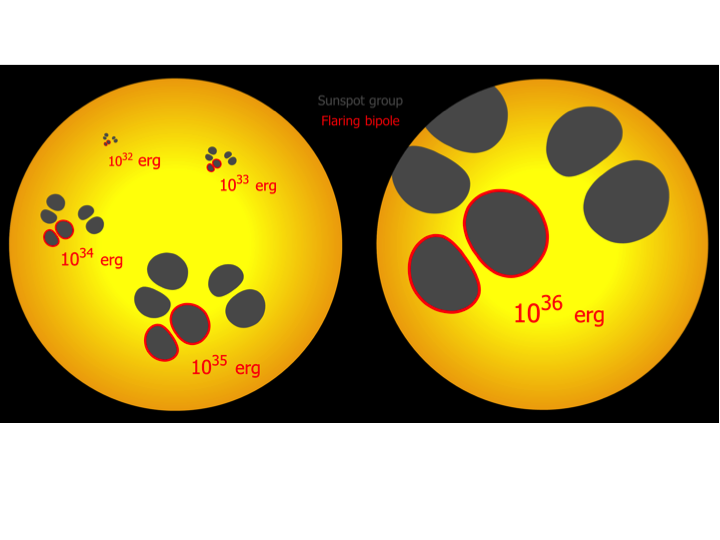}}
\centerline{\includegraphics[width=1\textwidth,clip=]{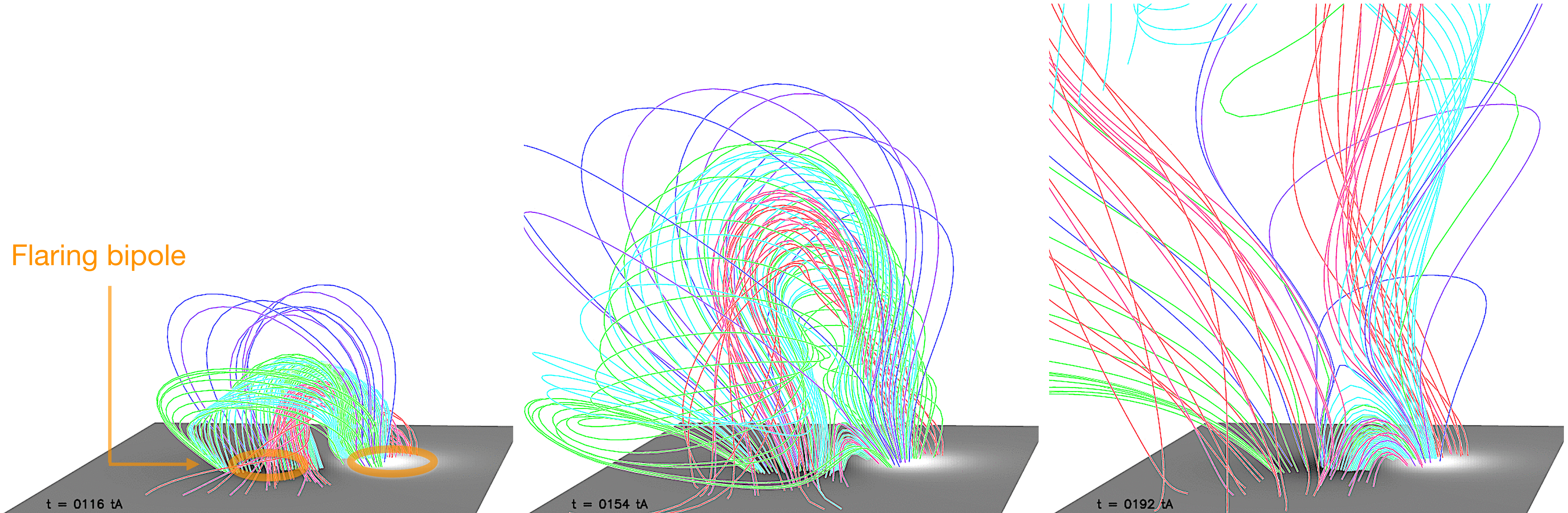}}
\caption{Sunspots and energetics of possible flares obtained by the OHM  MHD simulation; the main bipole spots which have to be considered active for the eruption are surrounded by red lines (adapted from \citet{Aulanier2013,Schmieder2018}). On the second row, the OHM magnetic field lines in a bipole are shown. The magnetic bipole is represented by two orange circles in the left panel. In the right panel, reconnection loops are depicted below the reconnection point (adapted from \citet{2016NatPh..12..998A}).}
\label{sunspot}
 \end{figure}
\begin{figure} 
\centerline{\includegraphics[width=0.45\textwidth,clip=]{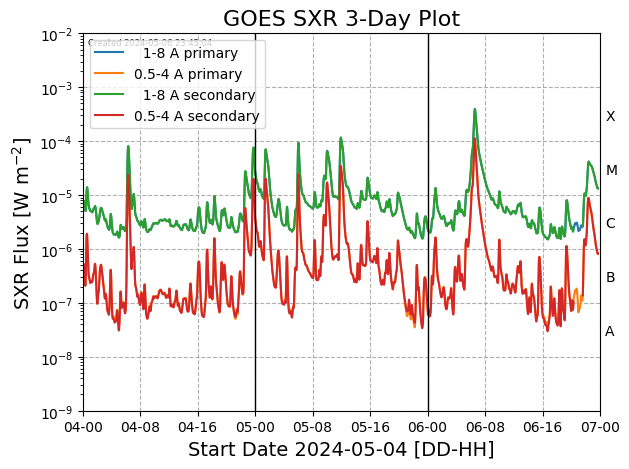}
\includegraphics[width=0.45\textwidth,clip=]{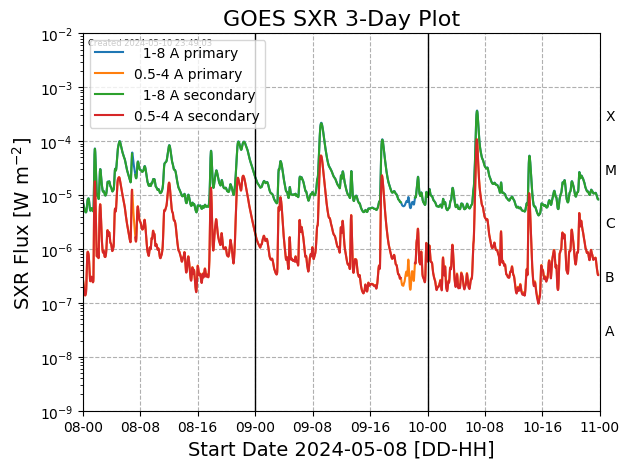}}
\caption{GOES data for the X-ray flares occurring in AR 13663 (left panel) and  AR 13664 (right panel).}
\label{fig:GOES}
 \end{figure}

\begin{figure} 
\centerline{
\includegraphics[width=0.95\textwidth,clip=]{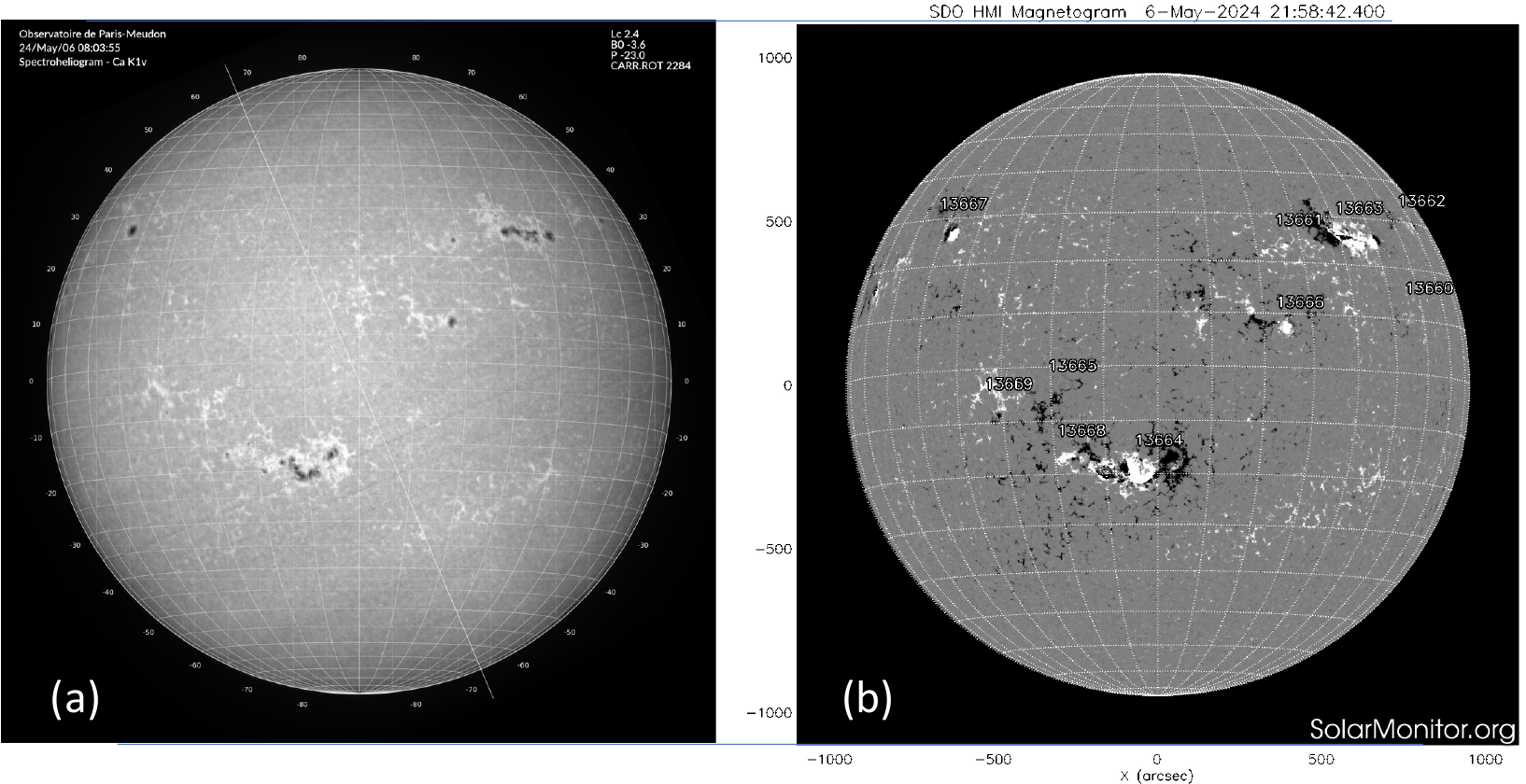}
}
\caption{Active regions AR13663 in the North and AR13664 in the South on May 6 2024, observed in Ca K line  by the Meudon spectroheliograph (a) and HMI (b).}
\label{fig:disk}
\end{figure}

This paper is focused on several simulation models: { OHM ("Observationally driven High order scheme Magnetohydrodynamic code"), a TMF (time-dependent magneto-friction) model to set up the initial non-potential magnetic field configuration combined with a zero-beta MHD model tracing the magnetic evolution of active regions, and EUHFORIA (''European heliospheric forecasting information asset''),} which aims to predict the solar wind and the evolution of CMEs in our space environment. The first step is to investigate the related solar activity, and then to detect the chain of events from the Sun to the Earth. We investigate three MHD simulations: the first simulation {(OHM)} is an ideal MHD simulation based on a bipole with rotation movements at the boundary. This mimics an active region with two sunspots submitted to photospheric motions (Section 2). { The second simulation (TMF) is constrained by observational data (successive vector magnetograms), i.e., a data-constrained simulation \citep{Guo2024}. These two simulations do not model the eruption itself, but only the pre-eruptive behaviour.} The stored free energy and the amount of helicity are computed during the evolution of the active region (Section 3). The last MHD simulation concerns the propagation and evolution of the CME in the heliosphere until it reaches the Earth, using the EUHFORIA \citep{Pomoell2018} (Section 4). The last two simulations concern the events that occurred in May 2024. In Section 5, we conclude that these three simulations are highly effective in predicting the likelihood of a large flare, a fast CME, and the arrival time at Earth of the disturbance.

\section{Prediction of extreme solar storms with MHD simulations}

As we noted in the introduction, extreme events consist of a chain of significant phenomena. The activity of the Sun is observed by the emergence of sunspots with strong magnetic fields and long-duration CMEs. It seems interesting to relate 
sunspot size and CME eruptive events.  For such a survey, the  {\it Observationally driven High order scheme Magnetohydrodynamic code} (OHM) \citep{Aulanier2005,Aulanier2010}  simulation has been used as an experiment to investigate the energy storage during the evolution of the bipole and to quantify the stored energy during the process of emergence \citep{Aulanier2013}. Such simulations can experiment with huge energetic events on the Sun and stars, e.g., large super flares, by exploring the characteristics of the spots in a large parameter space, ranging from the maximum size of solar sunspots observed in Meudon since 1907 (Figure~\ref{1972}) to the huge stellar spots with large flux. {We summarise below a few results obtained with OHM, which will be useful for the comparison with a data-constrained model in Section 3.}

The model consists of a bipole with two rotating spots, which is equivalent to creating a strong shear along the polarity inversion line with cancelling flux \citep{Aulanier2013}. The 3D numerical simulation solves the full MHD equations for the mass density, the fluid velocity $\mathbf{u}$, and the magnetic field $\mathbf{B}$ under the assumption of plasma $\beta =0$. The calculations were performed in non-dimensionalised
units, using $\mu = 1$.  The  CME is obtained by cancellation of the magnetic field at the solar surface (Figure \ref{sunspot} bottom panels). The details of the simulations and the results can be found in \citet{Aulanier2010,Zuccarello2015}.

The magnetic flux  $\Phi$ and the total flare energy $E$ are eventually given from the rescaling  as  in the following equations
\begin{eqnarray}
\noindent  \phi = 42 \left(\frac{B_z}{8T}\right)  \left(\frac{L^{\rm bipole}}{5m}\right)^2 Wb \label{eq1.1},  \\
\noindent E= \frac{40}{\mu}\left(\frac{B_z}{8T}\right)^2\left(\frac{L^{\rm bipole}}{5m}\right)^3   J . \label{eq1.2}
\end{eqnarray}

Here, $B$ denotes the strength of the magnetic field in the bipole (sunspot), and $L$ is the size of the bipole in micro solar hemisphere (MSH) units. The size of the concerned region has to be estimated using real observations. Does it correspond to the sunspots' full size or the flare's starting place, with the flare ribbons, where there is stress on the magnetic field lines?

With extreme values for the maximum flux ($\phi = 8 \times 10^{22}\;$Mx)  and large size of the flaring region ($L=100\;$Mm), corresponding to one third of the size of the largest 
sunspot group ever recorded back in April 1947, the maximum energy reaches $6 \times 10^{33}\;$erg. 
\citet{Cliver2022} argued that this number was overestimated for the flare energy, since this number actually corresponds to a magnetic energy drop, and some of it should go to { kinetic energy via the CME}, more than we calculated. 
So we should make a distinction between the stars with sunspots that can be larger than 300~Mm, where the energy
can be larger than $10^{35}\;$erg, and the Sun with a maximum sunspot group size like in 1947 or 1972, where the energy can not be larger than $6 \times 10^{33}\;$erg.  

In the past, with a younger Sun, it was perhaps similar to the young stars, which exhibit more activity. { Such large flares can occur} 
in the stars \citep{2012Natur.485..478M}.

\citet{Aulanier2013} give diagrams about the size of the sunspot and the estimated energy. To get superflares, an unrealistic size of sunspots and stress should be considered (Figure~\ref{sunspot} top panels).  

Therefore, we aim to investigate the two solar active regions of May 2024: NOAA AR 13663 and NOAA AR 13664, which produce a significant number of X-ray flares (Figures~\ref{fig:GOES} and \ref{fig:disk}). \citet{Hayakawa2025} claimed that from May 4 to 7, the main sunspot of AR 13664 grew in size from about 110 to 2700 MSH. They estimated the magnetic free energy to surpass $10^{33}\;$erg on May 7, triggering 12 X-class flares between May 8 and 15 \citep{Jarolim2024,Hayakawa2025}. These events initiated the strongest geomagnetic storm of solar cycle 26, with a Dst value of $-412\;$nT and an intense G5 geomagnetic storm \citep{Kwak2024}.   We could have expected a lower free energy according to our previous experiment, given the size of the spots when Figure~\ref{fig:disk} is compared with Figure~\ref{sunspot}.

\begin{figure} 
\centerline{\includegraphics[width=1.2\textwidth]{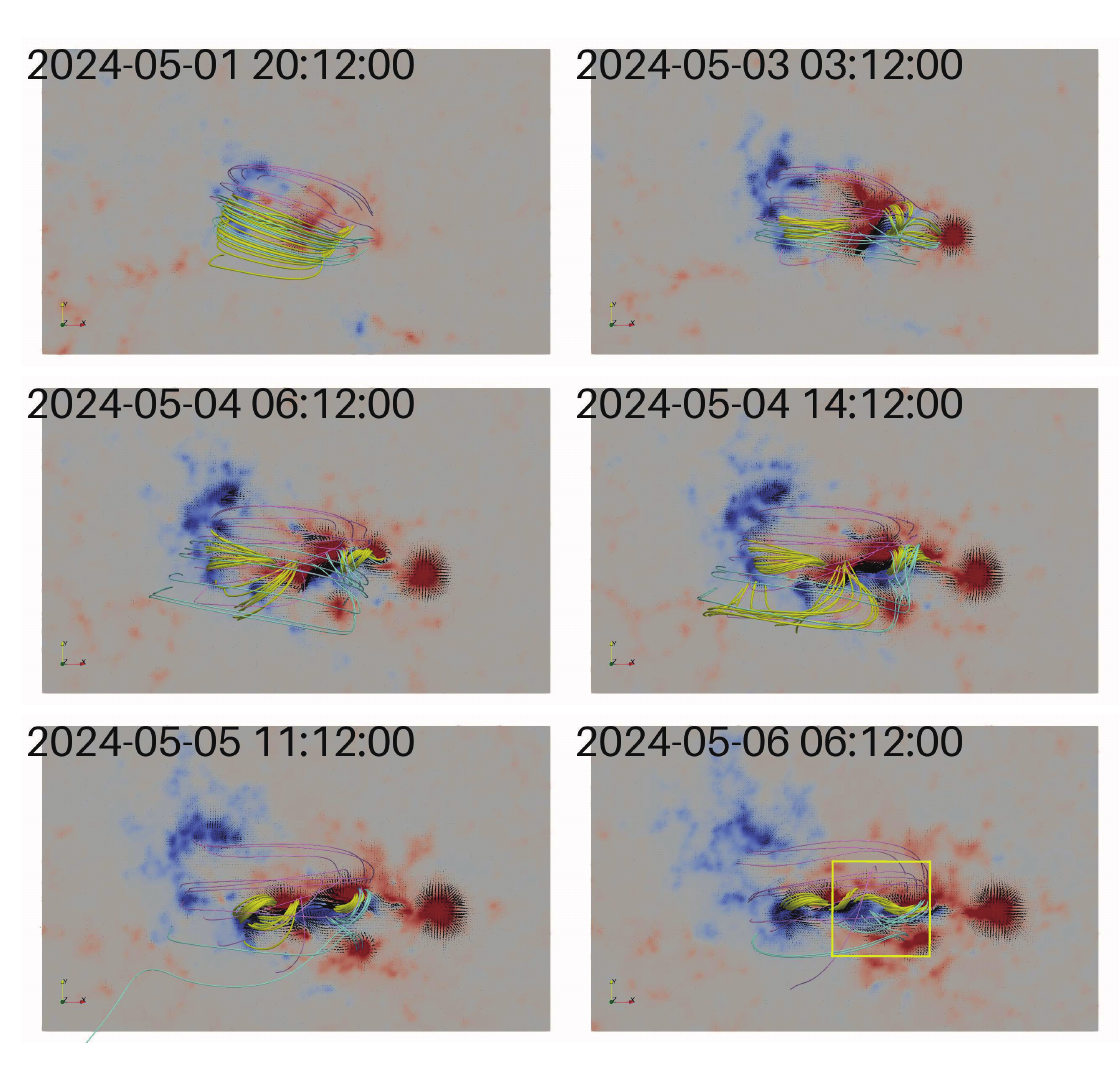}}
\caption{Magnetic-field evolution of AR 13663. The colour on the bottom surface represents the distribution of $B_z$  in the plane at 1 Mm height. In the lower right panel, we draw the box where the active region is smoothed as a bipole for comparison with the \citet{Aulanier2013} simulation.}\label{fig:evolution}
\end{figure}

\begin{figure} 
\centerline{\includegraphics[width=1.\textwidth,clip=]{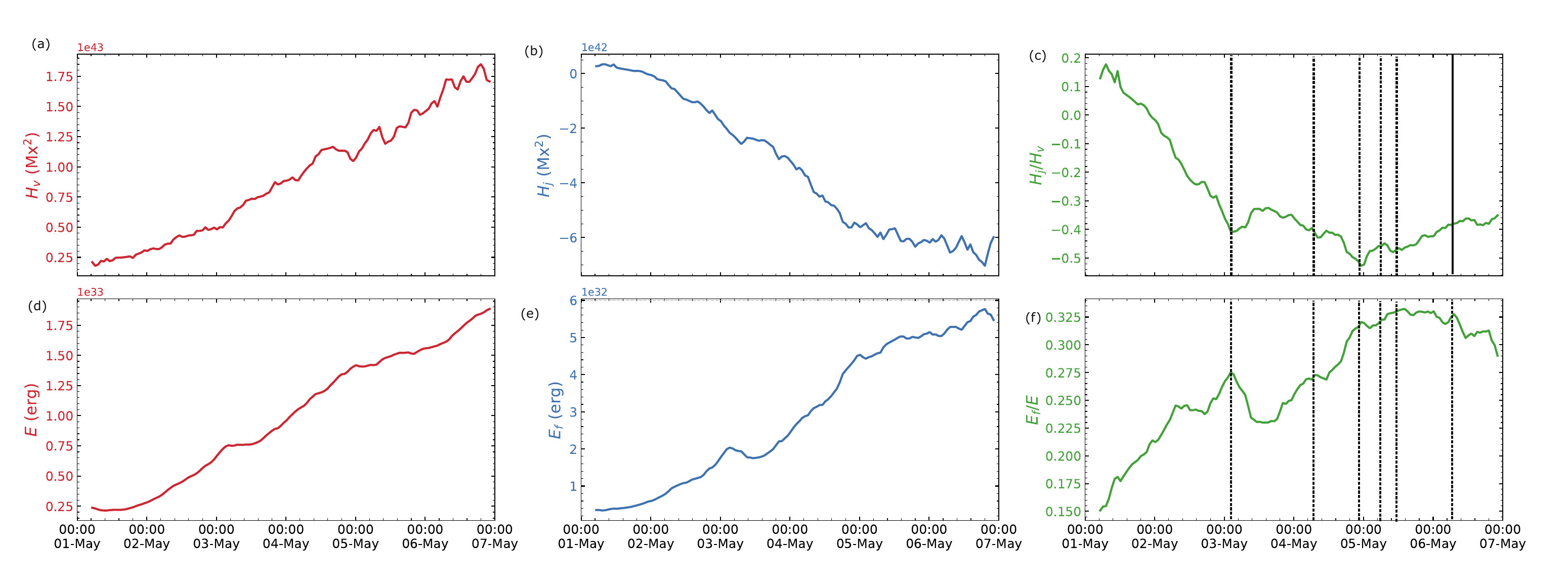}}
\caption{{Magnetic helicity: total helicity (a), current- carrying helicity (b), ratio (c). Magnetic energy: total (d), free magnetic energy (e), ratio (f)}.
The vertical red lines in panels (c, f) indicate the times of the flares. }
\label{fig:energy}
\end{figure}

\begin{figure} 
\centerline{\includegraphics[width=1.\textwidth,clip=]{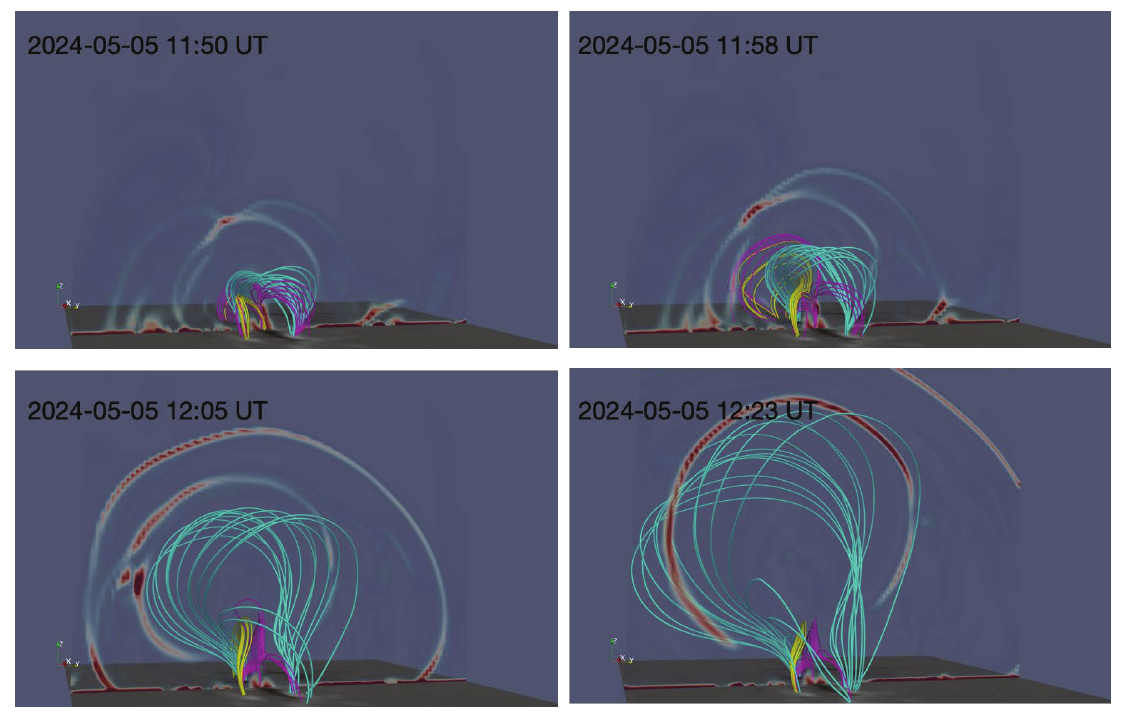}}
\caption{Evolution of the field lines during the eruption. The yellow lines represent the eruption of the flux rope, and the pink lines represent the current sheet below the flux rope.}
\label{fig:FR}
\end{figure}

\section{{ Data-driven simulation for the May 2024 events}}
Free magnetic energy evolution in an active region is a key determinant in predicting X-ray flares and significant geomagnetic events. For May 2024 events, various studies evaluate the maximum free energy before each X-ray flare, primarily using nonlinear force-free field extrapolations \citep{Jarolim2024,Hayakawa2025}.

Here, we perform a data-{driven} simulation to model the evolution of 3D coronal magnetic fields and the build-up of magnetic energy and helicity using similar methods and equations as in \citet{Guo2024}. The setup of our simulation includes two stages. Regarding the long-term evolution prior to the eruption, we employ the time-dependent magneto-frictional (TMF) model to simulate the quasi-static change in coronal magnetic fields. The governing equations are as follows:

\begin{eqnarray}
 && \frac{\partial \boldsymbol{B}}{\partial t} + \nabla \cdot(\boldsymbol{vB-Bv})=0,\label{eq1}\\
 && \boldsymbol{v}=\frac{1}{\nu}\ \frac{\boldsymbol{j \times B}}{ B^{2}},\label{eq2}\\
  && \nu= \frac{\nu_{0}}{1-e^{-z/L}},\label{eq3}
 \end{eqnarray}
where $\nu_0=10^{-15}\;$s\;cm$^{-2}$ is the viscous coefficient of the friction, $L=10\;$Mm is the decay spatial scale of the viscosity toward the boundaries and $\eta$ is the magnetic diffusivity, which are employed by \citet{Cheung2012} and \citet{Pomoell2019}. More details about the numerical scheme and the boundary conditions can be found in \citet{Guo2024}. In particular, the spatial size of the grids is twice the resolution of the original HMI instrument, resulting in a smoother image compared to the original magnetograms. Hereafter, we use the zero-$\beta$ MHD simulation to study the drastic eruption, in which the initial magnetic-field condition is provided by one snapshot of the TMF model before the flare. 
{The initial velocity of the zero-$\beta$ model is set to 0 throughout the computation domain, which is in accord with the force-free assumption for the magnetic state of the pre-eruptive flare.}
The governing equations of the zero-$\beta$ modeling are as follows:

\begin{eqnarray}
 && \frac{\partial \rho}{\partial t} +\nabla \cdot(\rho \boldsymbol{v})=0,\label{eq4}\\
 && \frac{\partial (\rho \boldsymbol{v})}{\partial t}+\nabla \cdot(\rho \boldsymbol{vv}+\frac{1}{2}B^{2}\boldsymbol{I}-\boldsymbol{\rm BB})=0,\label{eq5}\\
 && \frac{\partial \boldsymbol{B}}{\partial t} + \nabla \cdot(\boldsymbol{vB-Bv})=0.\label{eq6} 
\end{eqnarray}

{We adopt a three-step Runge-Kutta time stepper and the HLL Riemann solver with a three-order Cada limiter to numerically solve above equations.} Especially, we do not introduce explicit resistivity in the equations, as the numerical diffusion due to the difference scheme induces magnetic reconnection.

In this paper, we use NOAA {AR 13663} as an example and employ a data-driven simulation to reproduce its evolution. Figure~\ref{fig:evolution} shows the evolution of 3D magnetic field lines of AR 13663. The simulation clearly shows the response of coronal magnetic fields to the photospheric flows. From 20:12 UT on May 1 to 03:12 UT on May 3, new conjugate polarities emerge in the southwest, forming a quadrupole magnetic configuration. Hereafter, the newly emerged polarities collide with the north positive polarity to form a polarity inversion line (PIL), in which the magnetic fields above it become increasingly shearing and twisted. Finally, twisted flux ropes with negative helicity are formed, serving as the precursors of CMEs.

To further detail the evolution of this active region, we compute the magnetic energy and relative helicity based on 3D magnetic fields, with the following formula:

\begin{eqnarray}
  && E_{M}=\frac{1}{2\mu_{0}}\int \boldsymbol{B}^{2} dV,\label{eq12}\\
  && E_{free}=\frac{1}{2\mu_{0}}\int (\boldsymbol{B}^{2}-\boldsymbol{B_{p}}^{2}) dV,\label{eq13}\\
 && H_{R}=\int (\boldsymbol{A}+\boldsymbol{A_{p}})\cdot(\boldsymbol{B}-\boldsymbol{B_{p}}) dV=H_{J}+H_{pJ},\label{eq14}\\
    && H_{J}=\int (\boldsymbol{A}-\boldsymbol{A_{p}})\cdot(\boldsymbol{B}-\boldsymbol{B_{p}}) dV,\label{eq15}\\
  && H_{PJ}=2\int \boldsymbol{A_{p}}\cdot(\boldsymbol{B}-\boldsymbol{B_{p}}) dV,\label{eq16}\\
 \notag  
 \end{eqnarray}
where $E_M$ and $E_{free}$ represent total and free magnetic energy, respectively. $H_{R}$, $H_{J}$ and $H_{PJ}$ are the relative magnetic helicity, current-carrying helicity and mutual magnetic helicity, respectively. Especially, the vector potential is computed with the Devore gauge \citep{Devore2000a, Devore2000b, Valori2012}. The reference magnetic field (potential field) is computed from the Laplace equation constrained by six boundaries. Figure~\ref{fig:evolution} shows the evolution of helicity and magnetic energy. It is found that the magnetic energy and total relative helicity maintain an increasing trend all the time (Figure~\ref{fig:energy}). However, for the free energy and relative helicity, they reach a plateau at about May 5 with the values of $ -6\times 10^{42}\;$Mx$^{2}$ and $5\times 10^{32}\;$erg. This plateau appears to be related to the threshold number of the non-potential magnetic helicity ratio at the onset of solar eruptions, as identified in the study by \cite{Zuccarello2018}.  Figures~\ref{fig:evolution}c and f show the helicity and energy ratios, from which one can see the increase before the flare. As a result, our data-driven simulation suggests that the energy and helicity ratios are effective proxies for predicting solar flares. Hereafter, we model the solar flares that occurred on May 5 with the zero-$\beta$ MHD model shown in Figure~\ref{fig:FR}. Our simulation results exhibit the eruptive flux rope (yellow lines) and the driven shock. The role of torus instability and magnetic reconnection in triggering and accelerating the solar eruption will be discussed in a future paper.


{As an experiment for flare prediction, we use the formula of equations 1 and 2, and compare the results with those of the data-driven simulation. The formulas are valid for a simple bipole. This is not the case for this elongated region. Therefore, we must define a box where the active region can be smoothed as a bipole (Figure~\ref{fig:evolution} panel at 06:12 UT).
The flaring PIL is between the North with positive polarities and the South with negative polarities, so we choose a square-shaped box centred on this PIL, and the axis of the box follows the average PIL. The box size can be estimated from a compromise between the length of the flux rope along the PIL, excluding the rope footpoints, as in \citet{Aulanier2013} and the extension of the polarities across the PIL. The definition of the box is not straightforward in this region, as it is very elongated and not bipolar. 

To follow the "spirit" and model of \citet{Aulanier2013}, we simplify (approximate) the AR as a simple bipole, i.e., one polarity on each side of the flaring PIL, i.e., only the positive part in the North and only the negative part in the South. Hence, when considering the max(abs($B_{\rm zphoto}$)), we discard the negative polarity in the South.

By simplifying this AR, one actually supposes that there are no (relevant) connections between the positive and the negative polarities on the South side of the flaring AR; even though there are some connections, as shown in Figure~\ref{fig:evolution}. That is the limit of the approach to this AR proposed by \citet{Aulanier2013}.\\

We estimated the values for the drawn box in Figure~\ref{fig:evolution}.\\[3pt] 
\noindent(i) $B_{z, \rm max}$= ~2283.96 G;\\
(j) Area ~ $1985.94\;$Mm${^2}$, with a $L$ of the bipole of $63.02\;$Mm (f~2);\\
(k) Estimated flare energy (E): $5.2 \times 10^{32}\;$erg;\\
(l) Estimated  flux (phi): $1.9 \times 10^{22}\;$Mx.\\

{The flare energy and flux are estimated by Equations (1) and (2), respectively. The input parameters of these two formulas are derived from the simulation results.
}

This predicted flare energy matches the free energy obtained from the data-driven simulations, about $5.4 \times 10^{32}\;$erg. This indicates 
a good agreement exists between the two simulations.}

{  This comparison between the two simulations cannot be applied to all the ARs. AR 13664 deviates significantly from a bipolar sunspot, and the flares it produces originate
from different regions. As a result, it is not straightforward to estimate the flare energy based on this
model. For details of the magnetic fields of this active region, please refer to  \citet{Jarolim2024}.}

\section{Heliospheric propagation of CMEs using EUHFORIA}

 After observing a large, active region with a significant amount of free energy and predicting whether it will produce an X-ray flare and/or a CME, the next step is to simulate the CME's evolution and predict its arrival time at Earth. The partition of the energy, i.e., the part of the free energy that is released, is not well predictable. Therefore, it is helpful to have observations by a coronagraph to detect if any CMEs erupted. It is crucial to predict the arrival time of large flare-induced CMEs at Earth, as they may be individually geo-effective, or the interaction between CMEs could form complex, geo-effective ejecta. Hence, we propagate the CMEs, potentially associated with the X-class flare eruption, in the heliosphere to predict their arrival times at Earth. \\
 
 { We employ the "European heliospheric forecasting information asset" \citep[EUHFORIA,][]{Pomoell2018} to evolve the CMEs from 0.1 to 2.0~au. First, the solar wind is modelled with a modified Wang-Sheeley-Arge \citep[WSA;][]{Arge2003} coronal model, which uses an observed magnetogram as input. Then, the CMEs are injected at 0.1~au (the inner boundary of the heliospheric model of EUHFORIA) as time-dependent boundary conditions, using the cone CME model as implemented in \citet{Pomoell2018} and further improved by \citet{scolini2020}. } \\
 
\begin{figure}
    \centering
\includegraphics[width=\linewidth]{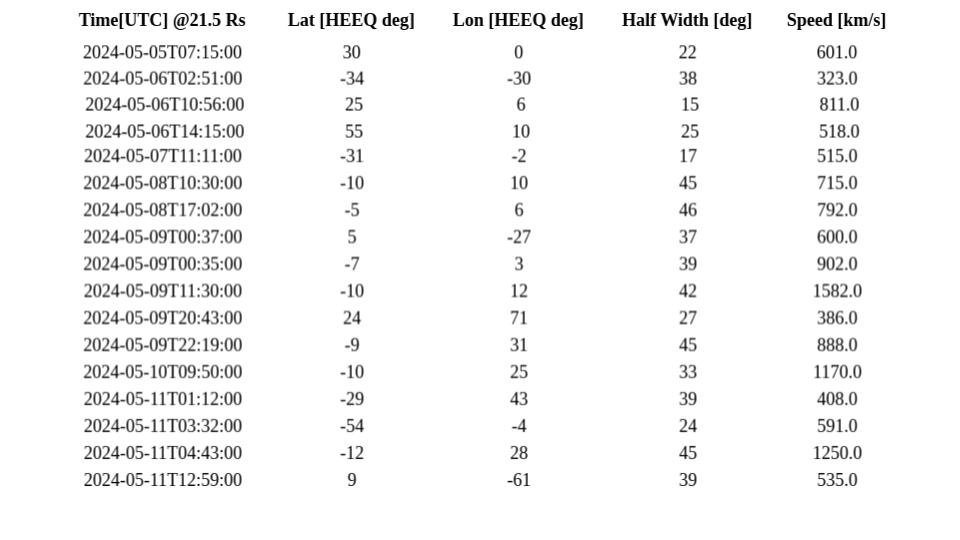}
    \caption{The combined list of CMEs considered in the EUHFORIA simulations and the corresponding parameters from the VSWMC portal.}
\label{fig:cme_params_vswmc}
\end{figure}

 In our study, we included the CMEs listed in the Virtual Space Weather Modelling Centre { (VSWMC - \url{https://swe.ssa.esa.int/gen_mod})} during the period May 5-11, 2024 (listed in Figure~\ref{fig:cme_params_vswmc}). In the VSWMC, the CME parameters can be retrieved automatically from the DONKI catalogue. We combined CME lists from two different VSWMC archived runs to form a complete list of all CMEs detected during that period. We have not performed a flare-CME correlation exercise. Instead, we just used the existing data for a quick demonstration of the arrival of the CMEs/merged CME-CME ejecta during the active periods of AR13663 and AR13664.\\

Once the CMEs are injected into the heliospheric domain of EUHFORIA, they are self-consistently evolved by solving the MHD equations until the CME arrives at 1~au or even beyond. This important chain of codes (WSA+EUHFORIA) is integrated into the VSWMC and validated by \citet{Rodriguez2024, Prete2024}. In our study, we have considered this chain using the simple cone CME model first to validate the arrival time of the CMEs at Earth. As the cone model is unmagnetized, we do not comment on the geo-effectiveness of the CMEs' impact. 

We performed two simulations with magnetograms of different dates: 
\begin{enumerate}
\item Run~1: Magnetogram from 2024-05-05 at 00:00 as activity began in AR13663 on this date. All CMEs in the VSWMC list are included (from May 5, 2024, to May 11, 2024). 
\item Run~2: Magnetogram from 2024-05-08 at 00:00 as major activity happened in AR13664 on this date. CMEs from May 8, 2024, to May 11, 2024, are included. 
\end{enumerate}
 The magnetograms are retrieved from the  GONG Air Force Data Assimilative Photospheric Flux
Transport Model (GONG-ADAPT) \citep{Arge2010,Hickmann2015}.  \\ 

\subsection*{Results}

\begin{figure} 
\centerline{
\includegraphics[width=1.25\textwidth,clip=,trim={0 5cm 0 4.5cm}]{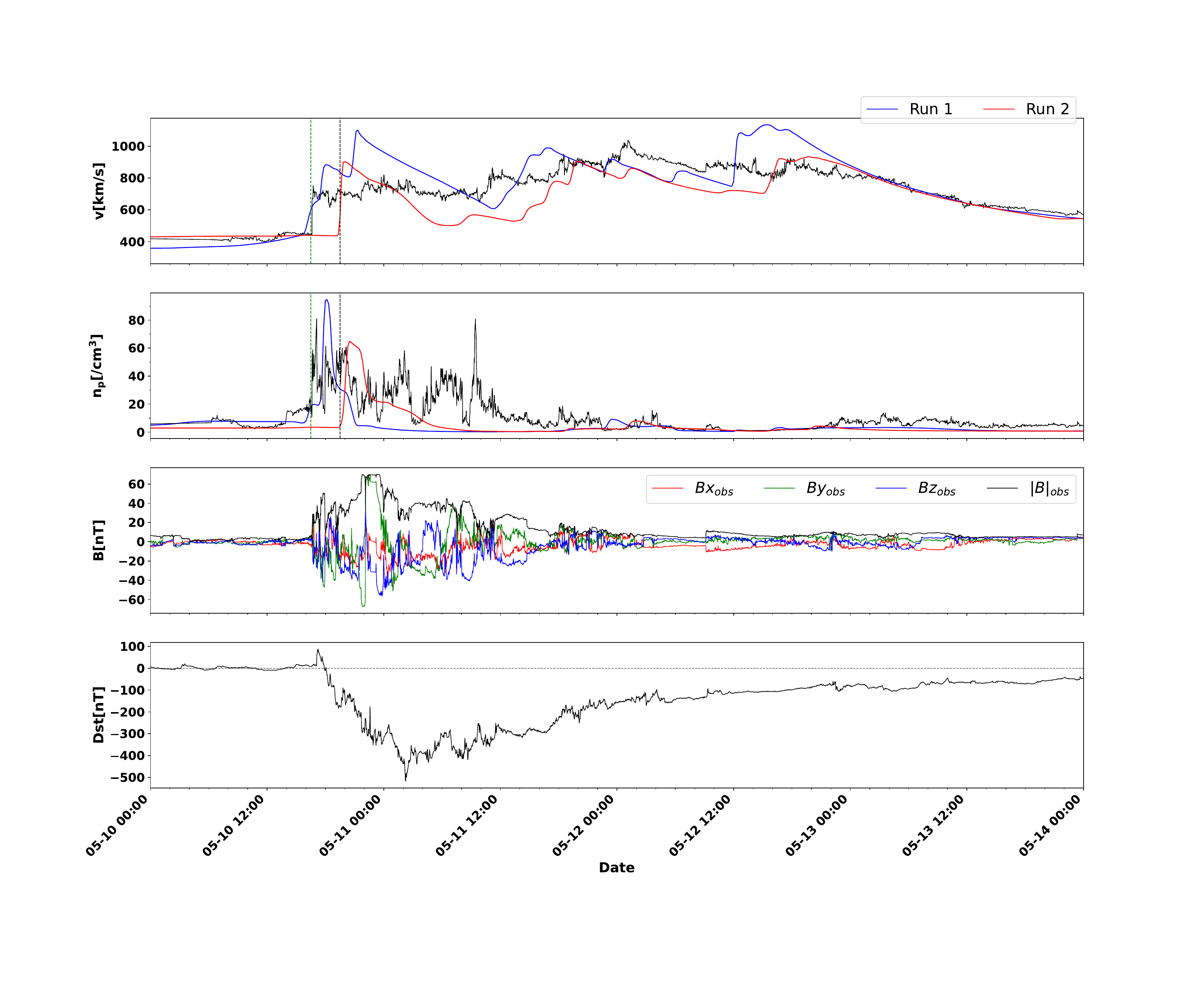}}
\caption{EUHFORIA simulation results with solar wind backgrounds generated using two different magnetograms. The in situ data measured by WIND is plotted in black. The first EUHFORIA run (Run 1) uses the magnetogram from 2024-05-05 at 00:00 { and consider all the CMEs listed in Figure 8 including the activity in AR13663 and in AR13664}, and the second run (Run 2) uses the magnetogram from 2024-05-08 00:00,    {the major activity is then concentrated in AR13664 with potential Earthward CMEs erupting}. The dashed vertical green and black lines in the speed and number density panels denote the observed arrival of the shock and the EUHFORIA-predicted shock from the second run. The third and fourth panels show the observed magnetic field and Dst value (no simulation results), respectively. { In the bottom panel the dashed curve indicates the zero value of Dst.}}
 \label{fig:cone_mags}
\end{figure}

The CME arrival time and proton number density predictions of the EUHFORIA simulations at Earth are plotted in Figure~\ref{fig:cone_mags}, compared to their corresponding measured values from the WIND spacecraft. The observed magnetic field profile and the Dst index are also provided for reference.

A shock signature is observed in the WIND speed profile on 2024-05-10 at 16:30. It is followed by a steady speed profile resembling a long-spanning sheath structure, which could result from merging multiple CMEs. In the case of Run~1, the first shock arrival aligns with the observations. However, the overall speed profile is overestimated. For Run~2 (magnetogram closer to the major eruptions), the arrival time is delayed by three hours. Nonetheless, although the profile is slightly underestimated, the overall profile qualitatively aligns better with the observations. The CMEs injected between 2024-05-05 and 2024-05-08 are not directed exactly towards Earth, but mostly are flank encounters at Earth (check Figure~\ref{fig:cme_params_vswmc} for their injection direction). These CMEs are still pre-conditioning the heliosphere, accelerating the CMEs launched on/after 2024-05-08, a critical aspect for accurate prediction.  

\begin{figure} 
\includegraphics[width=1\textwidth,clip=,trim={0 2.5cm 0 1cm}]{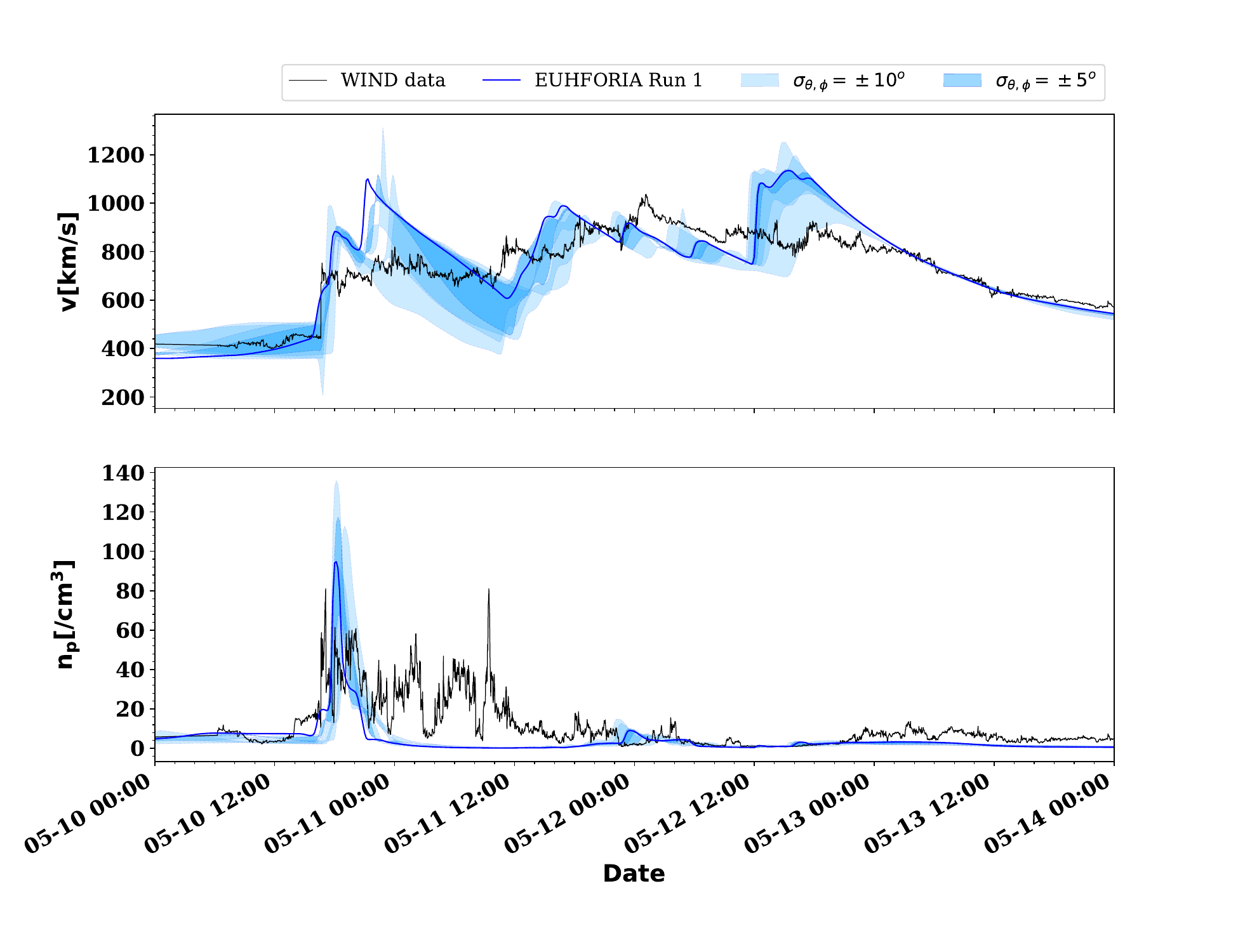} \\

\includegraphics[width=1\textwidth,clip=,trim={0 2.5cm 0 1cm}]{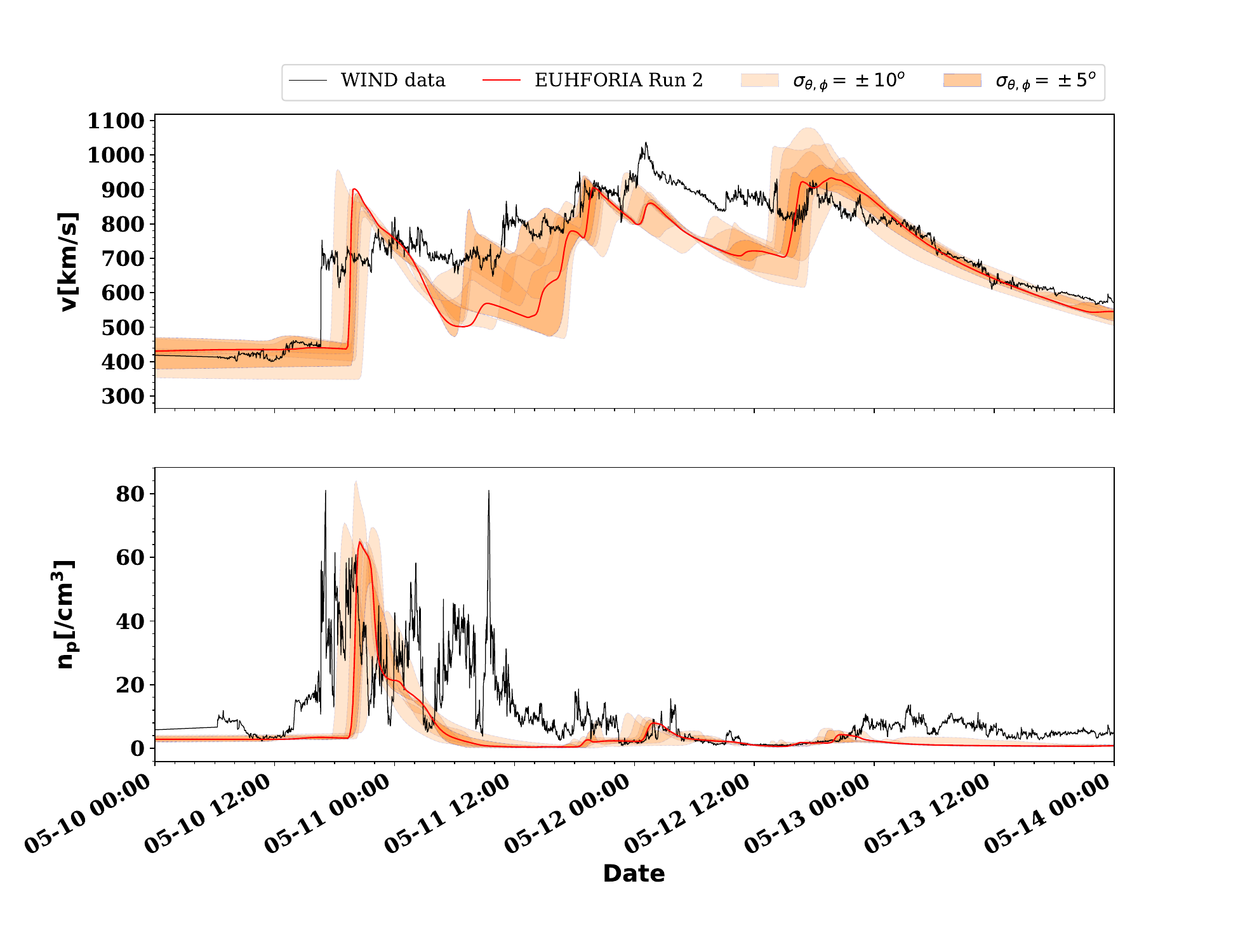}
\caption{The speed and number density profile of the best predicted EUHFORIA simulation (default VSWMC setup) plotted with the in situ data obtained by the WIND spacecraft for Run~1 (top) { using the magnetogram of May 5 2024} and Run~2 (bottom) { using the magnetogram of May 8 2024}. The light and dark shades around the EUHFORIA profile  {represent predictions at virtual satellites placed $5\deg$ and $10\deg$ away from Earth, providing some error bars on the direction of the CMEs. We still note a more accurate arrival time definition with Run 1.}} 
 \label{fig:cone_mag_best}
\end{figure}

In Figure~\ref{fig:cone_mag_best}, we have plotted both Run~1 and Run~2 speed and number density profiles at Earth along with the predictions at virtual spacecraft placed at locations $5\deg$ and $10\deg$ around Earth. Such plots serve as a proxy of the variability of the forecasts due to any errors in the initial CME input parameters. It is not straightforward to comment on whether Run~1 is better than Run~2 because it provides a more accurate shock arrival time estimate while overestimating the speed later. Through Run~2, we do realise that it is essential to consider all the CMEs in a given period as the complex CME-CME interactions dictate the overall dynamics and arrival time.

{The EUHFORIA model chain has been updated to enhance the realism of the simulations. The simple WSA-like coronal model has been upgraded to the full MHD model COCONUT \citep{Perri2022,Baratashvili2024}. Updating the input magnetograms enables the simulation of a dynamic solar Wind \citep{Wang2025_time}. The heliospheric wind model has also been made more efficient by applying radial grid stretching and adaptive mesh refinement (ICARUS) \citep{verbeke2022,Baratashvili2024,Baratashvili2025}. Moreover, different CME models have been introduced to take into account the internal magnetic field of the CMEs, which determines their geo-effectiveness: the spheromak \citep{Scolini2019}, the FRi3D \citep{Maharana2024}, the RBSL \citep{Guo2024_FR}, and several analytical torus and 'horseshoe' models   \citep{Linan2023,Linan2024,Maharana2024}. By combining the coronal and heliospheric model upgrades, an efficient global Sun-to-Earth model chain has been recently developed \citep{Linan2025,Baratashvili2024}. These models have all been validated using historic events, i.e., events that occurred in the past, so the measurements at 1~au and other locations are available. }

\section{Conclusion}

{In the present paper we  present three MHD models and their derived results  concerning  the amount of magnetic energy and magnetic helicity stored before huge events such as in May 2024.   We show with EUHFORIA how we are able to forecast the arrival time of a solar storm using a chain of MHD codes from the Sun to the Earth.  
 MHD simulations are very effective in predicting flares, CMEs and Geomagnetic storms. However, they all have their own limits. They should be developed and be closer to the observations. }
 The OHM simulation is relatively easy to use for predicting a superflare, but the sunspot size is not the only factor. It depends strongly on the {magnetic} stress of the region.  Nevertheless, we learn from this study that the main dipole in which the main pre-eruptive flux rope/filament/sigmoid is rooted before it erupts, drifts, expands, and eventually reconnects to more distant polarities. { This} limits the possible energy release to around 10$^{33}\;$erg for large sunspots. We must also consider that some of this energy can be released as kinetic energy through CMEs and accelerated particles \citep{Cliver2022,Warmuth2020}. The {data-constrained study of the May 2024 events} confirms the importance of the pre-eruptive flux rope.  The magnetic energy and the magnetic helicity reach a plateau with values around $-6\times 10^{42}\;$Mx$^{2}$ and $5\times 10^{32}\;$erg.
 { The data-driven} simulation is a step closer to reality. Many assumptions must be made to converge the results with the magneto-friction code, for example \citep{Schmieder2024}.  The computation is also time-consuming. 
{ Note  the  
 consistency between the data-initiated realistic simulation of an observed event 
and energy scalings from an idealised simulation of a bipolar eruption. The estimated magnetic free energy was comparable ($5.2 \times 10^{32}\;$erg) and significantly lower than the estimate given by an NLFF extrapolation approach \citep{Hayakawa2025}.  This shows the interest in coupling various numerical modelling approaches for space weather predictions, all inspired or directly driven by observations.}
EUHFORIA in the virtual space weather modelling centre is easy to use, but it does not consider the conditions for forming the flux rope, the cause of the CME, or the magnetic field of the ejecta. 
When the code update goes online, significant progress will be made in space weather forecasting. 
{ Based on multiple magnetograms with a time step of 5 minutes and a mesh containing 1.5~M cells, it will be possible to } capture the evolution of large-scale coronal structures and small-sized dipoles. Thus, this model is promising for accurately conducting real-time global coronal simulations of solar maximum, making it suitable for practical applications \citep{Wang2025}. { The artificial intelligence (AI)} is also applicable to predict magnetograms in advance for a few days \citep{Jeong2025}. All these codes are in development and need to be integrated into the Virtual Space Weather Modelling Centre.
Nowadays, AI  has emerged as a promising approach to enhance the predictive capabilities of space weather models 
 \citep{Guastavino2025}.  We will assess how efficient it is for predicting the CME arrival time.  

%

\subsection*{Acknowledgements}
\begin{acks}
Funded by the European Union. Views and opinions expressed are, however, those of the author(s) only and do not necessarily reflect those of the European Union or ERCEA. Neither the European Union nor the granting authority can be held responsible for them. This project (Open SESAME) has received funding under the Horizon Europe programme (ERC-AdG agreement No 101141362).

These results were also obtained in the framework of the projects C16/24/010 C1 project Internal Funds KU Leuven), G0B5823N and G002523N (WEAVE) (FWO-Vlaanderen), and 4000145223 (SIDC Data Exploitation (SIDEX2), ESA Prodex).

For the computations, we utilised the VSC (Flemish Supercomputer Centre) infrastructure, which was funded by the Hercules Foundation and the Flemish Government's EWI department. J.H.G. was supported by the China National Postdoctoral Program for Innovative Talents fellowship under Grant Number BX20240159.
{The work of BS and GA was supported by the Action Th\'ematique Soleil-Terre (ATST) of CNRS/INSU PN Astro, also funded by CNES, CEA, and ONERA. }
\end{acks}

%
%
%
%
%
%
%

%
%
\bibliographystyle{spr-mp-sola}
\bibliography{references}  

\begin{thebibliography}{60}
\ifx\bisbn     \undefined \def\bisbn  #1{ISBN #1}\fi
\ifx\binits    \undefined \def\binits#1{#1}\fi
\ifx\bauthor   \undefined \def\bauthor#1{#1}\fi
\ifx\batitle   \undefined \def\batitle#1{#1}\fi
\ifx\bjtitle   \undefined \def\bjtitle#1{\textit{#1}}\fi
\ifx\bvolume   \undefined \def\bvolume#1{\textbf{#1}}\fi
\ifx\byear     \undefined \def\byear#1{#1}\fi
\ifx\bissue    \undefined \def\bissue#1{#1}\fi
\ifx\bfpage    \undefined \def\bfpage#1{#1}\fi
\ifx\blpage    \undefined \def\blpage #1{#1}\fi
\ifx\burl      \undefined \def\burl#1{#1}\fi
\ifx\href      \undefined \def\href#1#2{#2}\fi
\ifx\betal     \undefined \def\betal{et al.}\fi
\ifx\bctitle   \undefined \def\bctitle#1{#1}\fi
\ifx\beditor   \undefined \def\beditor#1{#1}\fi
\ifx\bbtitle   \undefined \def\bbtitle#1{\textit{#1}}\fi
\ifx\bedition  \undefined \def\bedition#1{#1}\fi
\ifx\bseriesno \undefined \def\bseriesno#1{\textbf{#1}}\fi
\ifx\blocation \undefined \def\blocation#1{#1}\fi
\ifx\bsertitle \undefined \def\bsertitle#1{\textit{#1}}\fi
\ifx\bsnm      \undefined \def\bsnm#1{#1}\fi
\ifx\bsuffix   \undefined \def\bsuffix#1{#1}\fi
\ifx\bparticle \undefined \def\bparticle#1{#1}\fi
\ifx\barticle  \undefined \def\barticle#1{}\fi
\ifx\binstitute  \undefined \def\binstitute#1{#1}\fi
\ifx\bpublisher  \undefined \def\bpublisher#1{#1}\fi
\ifx\doiurl    \undefined \def\doiurl#1{\href{#1}{DOI}}\fi
\makeatletter
\def\safeHref#1#2#3{\in@{http}{#2}\ifin@\href{#2}{#3}\else\href{#1#2}{#3}\fi}
\makeatother
\ifx\adsurl    \undefined \def\adsurl#1{\safeHref{https://ui.adsabs.harvard.edu/abs/}{#1}{ADS}}\fi
\ifx\arxivurl  \undefined \def\arxivurl#1{\safeHref{http://arxiv.org/abs/}{#1}{arXiv}}\fi
\ifx\botherref \undefined \def\botherref#1{}\fi
\ifx\url       \undefined \def\url#1{#1}\fi
\ifx\bchapter  \undefined \def\bchapter#1{}\fi
\ifx\bbook     \undefined \def\bbook#1{}\fi
\ifx\bcomment  \undefined \def\bcomment#1{#1}\fi
\ifx\oauthor   \undefined \def\oauthor#1{#1}\fi
\ifx\citeauthoryear \undefined\def \citeauthoryear#1{#1}\fi
\def\endbibitem {}
\ifx\bconflocation  \undefined \def\bconflocation#1{#1} \fi

\bibitem[\protect\citeauthoryear{{Andrews}}{2003}]{Andrews2003}
\begin{barticle}
\bauthor{\bsnm{{Andrews}}, \binits{M.D.}}:
\byear{2003},
\batitle{{A Search for CMEs Associated with Big Flares}}.
\bjtitle{\solphys}
\bvolume{218},
\bfpage{261}.
\doiurl{https://doi.org/10.1023/B:SOLA.0000013039.69550.bf}.
\adsurl{2003SoPh..218..261A}.
\end{barticle}
\endbibitem

\bibitem[\protect\citeauthoryear{{Arge} et~al.}{2003}]{Arge2003}
\begin{bchapter}
\bauthor{\bsnm{{Arge}}, \binits{C.N.}},
\bauthor{\bsnm{{Odstrcil}}, \binits{D.}},
\bauthor{\bsnm{{Pizzo}}, \binits{V.J.}},
\bauthor{\bsnm{{Mayer}}, \binits{L.R.}}:
\byear{2003},
\bctitle{{Improved Method for Specifying Solar Wind Speed Near the Sun}}.
In: \beditor{\bsnm{{Velli}}, \binits{M.}},
\beditor{\bsnm{{Bruno}}, \binits{R.}},
\beditor{\bsnm{{Malara}}, \binits{F.}},
\beditor{\bsnm{{Bucci}}, \binits{B.}} (eds.)
\bbtitle{Solar Wind Ten},
\bsertitle{American Institute of Physics Conference Series}
\bseriesno{679},
\bfpage{190}.
\doiurl{https://doi.org/10.1063/1.1618574}.
\adsurl{2003AIPC..679..190A}.
\end{bchapter}
\endbibitem

\bibitem[\protect\citeauthoryear{{Arge} et~al.}{2010}]{Arge2010}
\begin{bchapter}
\bauthor{\bsnm{{Arge}}, \binits{C.N.}},
\bauthor{\bsnm{{Henney}}, \binits{C.J.}},
\bauthor{\bsnm{{Koller}}, \binits{J.}},
\bauthor{\bsnm{{Compeau}}, \binits{C.R.}},
\bauthor{\bsnm{{Young}}, \binits{S.}},
\bauthor{\bsnm{{MacKenzie}}, \binits{D.}},
\bauthor{\bsnm{{Fay}}, \binits{A.}},
\bauthor{\bsnm{{Harvey}}, \binits{J.W.}}:
\byear{2010},
\bctitle{{Air Force Data Assimilative Photospheric Flux Transport (ADAPT) Model}}.
In: \beditor{\bsnm{{Maksimovic}}, \binits{M.}},
\beditor{\bsnm{{Issautier}}, \binits{K.}},
\beditor{\bsnm{{Meyer-Vernet}}, \binits{N.}},
\beditor{\bsnm{{Moncuquet}}, \binits{M.}},
\beditor{\bsnm{{Pantellini}}, \binits{F.}} (eds.)
\bbtitle{Twelfth International Solar Wind Conference},
\bsertitle{American Institute of Physics Conference Series}
\bseriesno{1216},
\bpublisher{AIP},
\bfpage{343}.
\doiurl{https://doi.org/10.1063/1.3395870}.
\adsurl{2010AIPC.1216..343A}.
\end{bchapter}
\endbibitem

\bibitem[\protect\citeauthoryear{{Aulanier}}{2016}]{2016NatPh..12..998A}
\begin{barticle}
\bauthor{\bsnm{{Aulanier}}, \binits{G.}}:
\byear{2016},
\batitle{{Solar physics: When the tail wags the dog}}.
\bjtitle{Nature Physics}
\bvolume{12},
\bfpage{998}.
\doiurl{https://doi.org/10.1038/nphys3938}.
\adsurl{2016NatPh..12..998A}.
\end{barticle}
\endbibitem

\bibitem[\protect\citeauthoryear{{Aulanier}, {D{\'e}moulin}, and {Grappin}}{2005}]{Aulanier2005}
\begin{barticle}
\bauthor{\bsnm{{Aulanier}}, \binits{G.}},
\bauthor{\bsnm{{D{\'e}moulin}}, \binits{P.}},
\bauthor{\bsnm{{Grappin}}, \binits{R.}}:
\byear{2005},
\batitle{{Equilibrium and observational properties of line-tied twisted flux tubes}}.
\bjtitle{\aap}
\bvolume{430},
\bfpage{1067}.
\doiurl{https://doi.org/10.1051/0004-6361:20041519}.
\adsurl{2005A&A...430.1067A}.
\end{barticle}
\endbibitem

\bibitem[\protect\citeauthoryear{{Aulanier} et~al.}{2010}]{Aulanier2010}
\begin{barticle}
\bauthor{\bsnm{{Aulanier}}, \binits{G.}},
\bauthor{\bsnm{{T{\"o}r{\"o}k}}, \binits{T.}},
\bauthor{\bsnm{{D{\'e}moulin}}, \binits{P.}},
\bauthor{\bsnm{{DeLuca}}, \binits{E.E.}}:
\byear{2010},
\batitle{{Formation of Torus-Unstable Flux Ropes and Electric Currents in Erupting Sigmoids}}.
\bjtitle{\apj}
\bvolume{708},
\bfpage{314}.
\doiurl{https://doi.org/10.1088/0004-637X/708/1/314}.
\adsurl{2010ApJ...708..314A}.
\end{barticle}
\endbibitem

\bibitem[\protect\citeauthoryear{{Aulanier} et~al.}{2013}]{Aulanier2013}
\begin{barticle}
\bauthor{\bsnm{{Aulanier}}, \binits{G.}},
\bauthor{\bsnm{{D{\'e}moulin}}, \binits{P.}},
\bauthor{\bsnm{{Schrijver}}, \binits{C.J.}},
\bauthor{\bsnm{{Janvier}}, \binits{M.}},
\bauthor{\bsnm{{Pariat}}, \binits{E.}},
\bauthor{\bsnm{{Schmieder}}, \binits{B.}}:
\byear{2013},
\batitle{{The standard flare model in three dimensions. II. Upper limit on solar flare energy}}.
\bjtitle{\aap}
\bvolume{549},
\bfpage{A66}.
\doiurl{https://doi.org/10.1051/0004-6361/201220406}.
\adsurl{2013A&A...549A..66A}.
\end{barticle}
\endbibitem

\bibitem[\protect\citeauthoryear{{Baratashvili} et~al.}{2024}]{Baratashvili2024}
\begin{barticle}
\bauthor{\bsnm{{Baratashvili}}, \binits{T.}},
\bauthor{\bsnm{{Brchnelova}}, \binits{M.}},
\bauthor{\bsnm{{Linan}}, \binits{L.}},
\bauthor{\bsnm{{Lani}}, \binits{A.}},
\bauthor{\bsnm{{Poedts}}, \binits{S.}}:
\byear{2024},
\batitle{{The operationally ready full 3D magnetohydrodynamic model from the Sun to Earth: COCONUT+Icarus}}.
\bjtitle{\aap}
\bvolume{690},
\bfpage{A184}.
\doiurl{https://doi.org/10.1051/0004-6361/202449389}.
\adsurl{2024A&A...690A.184B}.
\end{barticle}
\endbibitem

\bibitem[\protect\citeauthoryear{{Baratashvili} et~al.}{2025}]{Baratashvili2025}
\begin{barticle}
\bauthor{\bsnm{{Baratashvili}}, \binits{T.}},
\bauthor{\bsnm{{Popescu Braileanu}}, \binits{B.}},
\bauthor{\bsnm{{Bacchini}}, \binits{F.}},
\bauthor{\bsnm{{Keppens}}, \binits{R.}},
\bauthor{\bsnm{{Poedts}}, \binits{S.}}:
\byear{2025},
\batitle{{Icarus 3.0: Dynamic heliosphere modelling}}.
\bjtitle{\aap}
\bvolume{694},
\bfpage{A306}.
\doiurl{https://doi.org/10.1051/0004-6361/202452705}.
\adsurl{2025A&A...694A.306B}.
\end{barticle}
\endbibitem

\bibitem[\protect\citeauthoryear{{Carmichael}}{1964}]{Carmichael1964}
\begin{bchapter}
\bauthor{\bsnm{{Carmichael}}, \binits{H.}}:
\byear{1964},
\bctitle{{A Process for Flares}}.
In: \beditor{\bsnm{{Hess}}, \binits{W.N.}} (ed.)
\bbtitle{NASA Special Publication}
\bseriesno{50},
\bfpage{451}.
\adsurl{1964NASSP..50..451C}.
\end{bchapter}
\endbibitem

\bibitem[\protect\citeauthoryear{{Chandra} et~al.}{2010}]{Chandra2010}
\begin{barticle}
\bauthor{\bsnm{{Chandra}}, \binits{R.}},
\bauthor{\bsnm{{Pariat}}, \binits{E.}},
\bauthor{\bsnm{{Schmieder}}, \binits{B.}},
\bauthor{\bsnm{{Mandrini}}, \binits{C.H.}},
\bauthor{\bsnm{{Uddin}}, \binits{W.}}:
\byear{2010},
\batitle{{How Can a Negative Magnetic Helicity Active Region Generate a Positive Helicity Magnetic Cloud?}}
\bjtitle{\solphys}
\bvolume{261},
\bfpage{127}.
\doiurl{https://doi.org/10.1007/s11207-009-9470-2}.
\adsurl{2010SoPh..261..127C}.
\end{barticle}
\endbibitem

\bibitem[\protect\citeauthoryear{{Cheung} and {Cameron}}{2012}]{Cheung2012}
\begin{barticle}
\bauthor{\bsnm{{Cheung}}, \binits{M.C.M.}},
\bauthor{\bsnm{{Cameron}}, \binits{R.H.}}:
\byear{2012},
\batitle{{Magnetohydrodynamics of the Weakly Ionized Solar Photosphere}}.
\bjtitle{\apj}
\bvolume{750},
\bfpage{6}.
\doiurl{https://doi.org/10.1088/0004-637X/750/1/6}.
\adsurl{2012ApJ...750....6C}.
\end{barticle}
\endbibitem

\bibitem[\protect\citeauthoryear{{Cliver} et~al.}{2022}]{Cliver2022}
\begin{barticle}
\bauthor{\bsnm{{Cliver}}, \binits{E.W.}},
\bauthor{\bsnm{{Schrijver}}, \binits{C.J.}},
\bauthor{\bsnm{{Shibata}}, \binits{K.}},
\bauthor{\bsnm{{Usoskin}}, \binits{I.G.}}:
\byear{2022},
\batitle{{Extreme solar events}}.
\bjtitle{Living Reviews in Solar Physics}
\bvolume{19},
\bfpage{2}.
\doiurl{https://doi.org/10.1007/s41116-022-00033-8}.
\adsurl{2022LRSP...19....2C}.
\end{barticle}
\endbibitem

\bibitem[\protect\citeauthoryear{{DeVore}}{2000}]{Devore2000a}
\begin{barticle}
\bauthor{\bsnm{{DeVore}}, \binits{C.R.}}:
\byear{2000},
\batitle{{Magnetic Helicity Generation by Solar Differential Rotation}}.
\bjtitle{\apj}
\bvolume{539},
\bfpage{944}.
\doiurl{https://doi.org/10.1086/309274}.
\adsurl{2000ApJ...539..944D}.
\end{barticle}
\endbibitem

\bibitem[\protect\citeauthoryear{{DeVore} and {Antiochos}}{2000}]{Devore2000b}
\begin{barticle}
\bauthor{\bsnm{{DeVore}}, \binits{C.R.}},
\bauthor{\bsnm{{Antiochos}}, \binits{S.K.}}:
\byear{2000},
\batitle{{Dynamical Formation and Stability of Helical Prominence Magnetic Fields}}.
\bjtitle{\apj}
\bvolume{539},
\bfpage{954}.
\doiurl{https://doi.org/10.1086/309275}.
\adsurl{2000ApJ...539..954D}.
\end{barticle}
\endbibitem

\bibitem[\protect\citeauthoryear{{Emslie} et~al.}{2004}]{Emslie2004}
\begin{barticle}
\bauthor{\bsnm{{Emslie}}, \binits{A.G.}},
\bauthor{\bsnm{{Kucharek}}, \binits{H.}},
\bauthor{\bsnm{{Dennis}}, \binits{B.R.}},
\bauthor{\bsnm{{Gopalswamy}}, \binits{N.}},
\bauthor{\bsnm{{Holman}}, \binits{G.D.}},
\bauthor{\bsnm{{Share}}, \binits{G.H.}},
\bauthor{\bsnm{{Vourlidas}}, \binits{A.}},
\bauthor{\bsnm{{Forbes}}, \binits{T.G.}},
\bauthor{\bsnm{{Gallagher}}, \binits{P.T.}},
\bauthor{\bsnm{{Mason}}, \binits{G.M.}},
\bauthor{\bsnm{{Metcalf}}, \binits{T.R.}},
\bauthor{\bsnm{{Mewaldt}}, \binits{R.A.}},
\bauthor{\bsnm{{Murphy}}, \binits{R.J.}},
\bauthor{\bsnm{{Schwartz}}, \binits{R.A.}},
\bauthor{\bsnm{{Zurbuchen}}, \binits{T.H.}}:
\byear{2004},
\batitle{{Energy partition in two solar flare/CME events}}.
\bjtitle{Journal of Geophysical Research (Space Physics)}
\bvolume{109},
\bfpage{A10104}.
\doiurl{https://doi.org/10.1029/2004JA010571}.
\adsurl{2004JGRA..10910104E}.
\end{barticle}
\endbibitem

\bibitem[\protect\citeauthoryear{{Forbes} et~al.}{2006}]{Forbes2006}
\begin{barticle}
\bauthor{\bsnm{{Forbes}}, \binits{T.G.}},
\bauthor{\bsnm{{Linker}}, \binits{J.A.}},
\bauthor{\bsnm{{Chen}}, \binits{J.}},
\bauthor{\bsnm{{Cid}}, \binits{C.}},
\bauthor{\bsnm{{K{\'o}ta}}, \binits{J.}},
\bauthor{\bsnm{{Lee}}, \binits{M.A.}},
\bauthor{\bsnm{{Mann}}, \binits{G.}},
\bauthor{\bsnm{{Miki{\'c}}}, \binits{Z.}},
\bauthor{\bsnm{{Potgieter}}, \binits{M.S.}},
\bauthor{\bsnm{{Schmidt}}, \binits{J.M.}},
\bauthor{\bsnm{{Siscoe}}, \binits{G.L.}},
\bauthor{\bsnm{{Vainio}}, \binits{R.}},
\bauthor{\bsnm{{Antiochos}}, \binits{S.K.}},
\bauthor{\bsnm{{Riley}}, \binits{P.}}:
\byear{2006},
\batitle{{CME Theory and Models}}.
\bjtitle{\ssr}
\bvolume{123},
\bfpage{251}.
\doiurl{https://doi.org/10.1007/s11214-006-9019-8}.
\adsurl{2006SSRv..123..251F}.
\end{barticle}
\endbibitem

\bibitem[\protect\citeauthoryear{{Gosling}}{1994}]{Gosling1994}
\begin{barticle}
\bauthor{\bsnm{{Gosling}}, \binits{J.T.}}:
\byear{1994},
\batitle{{Correction to ``The solar flare myth''}}.
\bjtitle{\jgr}
\bvolume{99},
\bfpage{4259}.
\doiurl{https://doi.org/10.1029/94JA00015}.
\adsurl{1994JGR....99.4259G}.
\end{barticle}
\endbibitem

\bibitem[\protect\citeauthoryear{{Guastavino} et~al.}{2025}]{Guastavino2025}
\begin{barticle}
\bauthor{\bsnm{{Guastavino}}, \binits{S.}},
\bauthor{\bsnm{{Legnaro}}, \binits{E.}},
\bauthor{\bsnm{{Massone}}, \binits{A.M.}},
\bauthor{\bsnm{{Piana}}, \binits{M.}}:
\byear{2025},
\batitle{{Artificial Intelligence for the Characterization of the 2024 May Superstorm: Active Region Classification, Flare Forecasting, and Geomagnetic Storm Prediction}}.
\bjtitle{\apj}
\bvolume{985},
\bfpage{53}.
\doiurl{https://doi.org/10.3847/1538-4357/adcd62}.
\adsurl{2025ApJ...985...53G}.
\end{barticle}
\endbibitem

\bibitem[\protect\citeauthoryear{{Guo} et~al.}{2024a}]{Guo2024}
\begin{barticle}
\bauthor{\bsnm{{Guo}}, \binits{J.H.}},
\bauthor{\bsnm{{Ni}}, \binits{Y.W.}},
\bauthor{\bsnm{{Guo}}, \binits{Y.}},
\bauthor{\bsnm{{Xia}}, \binits{C.}},
\bauthor{\bsnm{{Schmieder}}, \binits{B.}},
\bauthor{\bsnm{{Poedts}}, \binits{S.}},
\bauthor{\bsnm{{Zhong}}, \binits{Z.}},
\bauthor{\bsnm{{Zhou}}, \binits{Y.H.}},
\bauthor{\bsnm{{Yu}}, \binits{F.}},
\bauthor{\bsnm{{Chen}}, \binits{P.F.}}:
\byear{2024}a,
\batitle{{Data-driven Modeling of a Coronal Magnetic Flux Rope: From Birth to Death}}.
\bjtitle{\apj}
\bvolume{961},
\bfpage{140}.
\doiurl{https://doi.org/10.3847/1538-4357/ad088d}.
\adsurl{2024ApJ...961..140G}.
\end{barticle}
\endbibitem

\bibitem[\protect\citeauthoryear{{Guo} et~al.}{2024b}]{Guo2024_FR}
\begin{barticle}
\bauthor{\bsnm{{Guo}}, \binits{J.H.}},
\bauthor{\bsnm{{Linan}}, \binits{L.}},
\bauthor{\bsnm{{Poedts}}, \binits{S.}},
\bauthor{\bsnm{{Guo}}, \binits{Y.}},
\bauthor{\bsnm{{Schmieder}}, \binits{B.}},
\bauthor{\bsnm{{Lani}}, \binits{A.}},
\bauthor{\bsnm{{Ni}}, \binits{Y.W.}},
\bauthor{\bsnm{{Brchnelova}}, \binits{M.}},
\bauthor{\bsnm{{Perri}}, \binits{B.}},
\bauthor{\bsnm{{Baratashvili}}, \binits{T.}},
\bauthor{\bsnm{{Li}}, \binits{S.T.}},
\bauthor{\bsnm{{Chen}}, \binits{P.F.}}:
\byear{2024}b,
\batitle{{Dependence of coronal mass ejections on the morphology and toroidal flux of their source magnetic flux ropes}}.
\bjtitle{\aap}
\bvolume{690},
\bfpage{A189}.
\doiurl{https://doi.org/10.1051/0004-6361/202449731}.
\adsurl{2024A&A...690A.189G}.
\end{barticle}
\endbibitem

\bibitem[\protect\citeauthoryear{{Hayakawa}}{2025}]{Hayakawa2025}
\begin{botherref}
\oauthor{\bsnm{{Hayakawa}}, \binits{H.}}:
2025,
{The long-term solar variability, as reconstructed from historical sources: Several case studies in the 17th -- 18th centuries}.
\textit{arXiv e-prints},
arXiv:2502.14665.
\doiurl{https://doi.org/10.48550/arXiv.2502.14665}.
\adsurl{2025arXiv250214665H}.
\end{botherref}
\endbibitem

\bibitem[\protect\citeauthoryear{{Hickmann} et~al.}{2015}]{Hickmann2015}
\begin{barticle}
\bauthor{\bsnm{{Hickmann}}, \binits{K.S.}},
\bauthor{\bsnm{{Godinez}}, \binits{H.C.}},
\bauthor{\bsnm{{Henney}}, \binits{C.J.}},
\bauthor{\bsnm{{Arge}}, \binits{C.N.}}:
\byear{2015},
\batitle{{Data Assimilation in the ADAPT Photospheric Flux Transport Model}}.
\bjtitle{\solphys}
\bvolume{290},
\bfpage{1105}.
\doiurl{https://doi.org/10.1007/s11207-015-0666-3}.
\adsurl{2015SoPh..290.1105H}.
\end{barticle}
\endbibitem

\bibitem[\protect\citeauthoryear{{Hirayama}}{1974}]{Hirayama1974}
\begin{barticle}
\bauthor{\bsnm{{Hirayama}}, \binits{T.}}:
\byear{1974},
\batitle{{Theoretical Model of Flares and Prominences. I: Evaporating Flare Model}}.
\bjtitle{\solphys}
\bvolume{34},
\bfpage{323}.
\doiurl{https://doi.org/10.1007/BF00153671}.
\adsurl{1974SoPh...34..323H}.
\end{barticle}
\endbibitem

\bibitem[\protect\citeauthoryear{{Jarolim} et~al.}{2024}]{Jarolim2024}
\begin{barticle}
\bauthor{\bsnm{{Jarolim}}, \binits{R.}},
\bauthor{\bsnm{{Veronig}}, \binits{A.M.}},
\bauthor{\bsnm{{Purkhart}}, \binits{S.}},
\bauthor{\bsnm{{Zhang}}, \binits{P.}},
\bauthor{\bsnm{{Rempel}}, \binits{M.}}:
\byear{2024},
\batitle{{Magnetic Field Evolution of the Solar Active Region 13664}}.
\bjtitle{\apjl}
\bvolume{976},
\bfpage{L12}.
\doiurl{https://doi.org/10.3847/2041-8213/ad8914}.
\adsurl{2024ApJ...976L..12J}.
\end{barticle}
\endbibitem

\bibitem[\protect\citeauthoryear{{Jeong} et~al.}{2025}]{Jeong2025}
\begin{barticle}
\bauthor{\bsnm{{Jeong}}, \binits{H.-J.}},
\bauthor{\bsnm{{Jeon}}, \binits{M.}},
\bauthor{\bsnm{{Kim}}, \binits{D.}},
\bauthor{\bsnm{{Kim}}, \binits{Y.}},
\bauthor{\bsnm{{Baek}}, \binits{J.-H.}},
\bauthor{\bsnm{{Moon}}, \binits{Y.-J.}},
\bauthor{\bsnm{{Choi}}, \binits{S.}}:
\byear{2025},
\batitle{{Prediction of the Next Solar Rotation Synoptic Maps Using an Artificial Intelligence{\textendash}based Surface Flux Transport Model}}.
\bjtitle{Astophysics Journal Supplement}
\bvolume{278},
\bfpage{5}.
\doiurl{https://doi.org/10.3847/1538-4365/adc447}.
\adsurl{2025ApJS..278....5J}.
\end{barticle}
\endbibitem

\bibitem[\protect\citeauthoryear{{Kahler}}{1982}]{Kahler1982}
\begin{barticle}
\bauthor{\bsnm{{Kahler}}, \binits{S.W.}}:
\byear{1982},
\batitle{{The role of the big flare syndrome in correlations of solar energetic proton fluxes and associated microwave burst parameters}}.
\bjtitle{\jgr}
\bvolume{87},
\bfpage{3439}.
\doiurl{https://doi.org/10.1029/JA087iA05p03439}.
\adsurl{1982JGR....87.3439K}.
\end{barticle}
\endbibitem

\bibitem[\protect\citeauthoryear{{Kopp} and {Pneuman}}{1976}]{Kopp1976}
\begin{barticle}
\bauthor{\bsnm{{Kopp}}, \binits{R.A.}},
\bauthor{\bsnm{{Pneuman}}, \binits{G.W.}}:
\byear{1976},
\batitle{{Magnetic reconnection in the corona and the loop prominence phenomenon.}}
\bjtitle{\solphys}
\bvolume{50},
\bfpage{85}.
\doiurl{https://doi.org/10.1007/BF00206193}.
\adsurl{1976SoPh...50...85K}.
\end{barticle}
\endbibitem

\bibitem[\protect\citeauthoryear{{Kwak} et~al.}{2024}]{Kwak2024}
\begin{barticle}
\bauthor{\bsnm{{Kwak}}, \binits{Y.-S.}},
\bauthor{\bsnm{{Kim}}, \binits{J.-H.}},
\bauthor{\bsnm{{Kim}}, \binits{S.}},
\bauthor{\bsnm{{Miyashita}}, \binits{Y.}},
\bauthor{\bsnm{{Yang}}, \binits{T.}},
\bauthor{\bsnm{{Park}}, \binits{S.-H.}},
\bauthor{\bsnm{{Lim}}, \binits{E.-K.}},
\bauthor{\bsnm{{Jung}}, \binits{J.}},
\bauthor{\bsnm{{Kam}}, \binits{H.}},
\bauthor{\bsnm{{Lee}}, \binits{J.}},
\bauthor{\bsnm{{Lee}}, \binits{H.}},
\bauthor{\bsnm{{Yoo}}, \binits{J.-H.}},
\bauthor{\bsnm{{Lee}}, \binits{H.}},
\bauthor{\bsnm{{Kwon}}, \binits{R.-Y.}},
\bauthor{\bsnm{{Seough}}, \binits{J.}},
\bauthor{\bsnm{{Nam}}, \binits{U.-W.}},
\bauthor{\bsnm{{Lee}}, \binits{W.K.}},
\bauthor{\bsnm{{Hong}}, \binits{J.}},
\bauthor{\bsnm{{Sohn}}, \binits{J.}},
\bauthor{\bsnm{{Kwak}}, \binits{J.}},
\bauthor{\bsnm{{Kwak}}, \binits{H.}},
\bauthor{\bsnm{{Kim}}, \binits{R.-S.}},
\bauthor{\bsnm{{Kim}}, \binits{Y.-H.}},
\bauthor{\bsnm{{Cho}}, \binits{K.-S.}},
\bauthor{\bsnm{{Park}}, \binits{J.}},
\bauthor{\bsnm{{Lee}}, \binits{J.}},
\bauthor{\bsnm{{Nguyen}}, \binits{H.N.H.}},
\bauthor{\bsnm{{Talha}}, \binits{M.}}:
\byear{2024},
\batitle{{Observational Overview of the May 2024 G5-Level Geomagnetic Storm: From Solar Eruptions to Terrestrial Consequences}}.
\bjtitle{Journal of Astronomy and Space Sciences}
\bvolume{41},
\bfpage{171}.
\doiurl{https://doi.org/10.5140/JASS.2024.41.3.171}.
\adsurl{2024JASS...41..171K}.
\end{barticle}
\endbibitem

\bibitem[\protect\citeauthoryear{{Lin} and {Forbes}}{2000}]{Lin2000}
\begin{barticle}
\bauthor{\bsnm{{Lin}}, \binits{J.}},
\bauthor{\bsnm{{Forbes}}, \binits{T.G.}}:
\byear{2000},
\batitle{{Effects of reconnection on the coronal mass ejection process}}.
\bjtitle{\jgr}
\bvolume{105},
\bfpage{2375}.
\doiurl{https://doi.org/10.1029/1999JA900477}.
\adsurl{2000JGR...105.2375L}.
\end{barticle}
\endbibitem

\bibitem[\protect\citeauthoryear{{Linan} et~al.}{2023}]{Linan2023}
\begin{barticle}
\bauthor{\bsnm{{Linan}}, \binits{L.}},
\bauthor{\bsnm{{Regnault}}, \binits{F.}},
\bauthor{\bsnm{{Perri}}, \binits{B.}},
\bauthor{\bsnm{{Brchnelova}}, \binits{M.}},
\bauthor{\bsnm{{Kuzma}}, \binits{B.}},
\bauthor{\bsnm{{Lani}}, \binits{A.}},
\bauthor{\bsnm{{Poedts}}, \binits{S.}},
\bauthor{\bsnm{{Schmieder}}, \binits{B.}}:
\byear{2023},
\batitle{{Self-consistent propagation of flux ropes in realistic coronal simulations}}.
\bjtitle{\aap}
\bvolume{675},
\bfpage{A101}.
\doiurl{https://doi.org/10.1051/0004-6361/202346235}.
\adsurl{2023A&A...675A.101L}.
\end{barticle}
\endbibitem

\bibitem[\protect\citeauthoryear{{Linan} et~al.}{2024}]{Linan2024}
\begin{barticle}
\bauthor{\bsnm{{Linan}}, \binits{L.}},
\bauthor{\bsnm{{Maharana}}, \binits{A.}},
\bauthor{\bsnm{{Poedts}}, \binits{S.}},
\bauthor{\bsnm{{Schmieder}}, \binits{B.}},
\bauthor{\bsnm{{Keppens}}, \binits{R.}}:
\byear{2024},
\batitle{{Toroidal Miller-Turner and Soloviev coronal mass ejection models in EUHFORIA. I. Implementation}}.
\bjtitle{\aap}
\bvolume{681},
\bfpage{A103}.
\doiurl{https://doi.org/10.1051/0004-6361/202347794}.
\adsurl{2024A&A...681A.103L}.
\end{barticle}
\endbibitem

\bibitem[\protect\citeauthoryear{{Linan} et~al.}{2025}]{Linan2025}
\begin{barticle}
\bauthor{\bsnm{{Linan}}, \binits{L.}},
\bauthor{\bsnm{{Baratashvili}}, \binits{T.}},
\bauthor{\bsnm{{Lani}}, \binits{A.}},
\bauthor{\bsnm{{Schmieder}}, \binits{B.}},
\bauthor{\bsnm{{Brchnelova}}, \binits{M.}},
\bauthor{\bsnm{{Guo}}, \binits{J.H.}},
\bauthor{\bsnm{{Poedts}}, \binits{S.}}:
\byear{2025},
\batitle{{Coronal mass ejection propagation in the dynamically coupled space weather tool: COCONUT + EUHFORIA}}.
\bjtitle{\aap}
\bvolume{693},
\bfpage{A229}.
\doiurl{https://doi.org/10.1051/0004-6361/202451854}.
\adsurl{2025A&A...693A.229L}.
\end{barticle}
\endbibitem

\bibitem[\protect\citeauthoryear{{Maehara} et~al.}{2012}]{2012Natur.485..478M}
\begin{barticle}
\bauthor{\bsnm{{Maehara}}, \binits{H.}},
\bauthor{\bsnm{{Shibayama}}, \binits{T.}},
\bauthor{\bsnm{{Notsu}}, \binits{S.}},
\bauthor{\bsnm{{Notsu}}, \binits{Y.}},
\bauthor{\bsnm{{Nagao}}, \binits{T.}},
\bauthor{\bsnm{{Kusaba}}, \binits{S.}},
\bauthor{\bsnm{{Honda}}, \binits{S.}},
\bauthor{\bsnm{{Nogami}}, \binits{D.}},
\bauthor{\bsnm{{Shibata}}, \binits{K.}}:
\byear{2012},
\batitle{{Superflares on solar-type stars}}.
\bjtitle{\nat}
\bvolume{485},
\bfpage{478}.
\doiurl{https://doi.org/10.1038/nature11063}.
\adsurl{2012Natur.485..478M}.
\end{barticle}
\endbibitem

\bibitem[\protect\citeauthoryear{{Maharana} et~al.}{2024}]{Maharana2024}
\begin{barticle}
\bauthor{\bsnm{{Maharana}}, \binits{A.}},
\bauthor{\bsnm{{Linan}}, \binits{L.}},
\bauthor{\bsnm{{Poedts}}, \binits{S.}},
\bauthor{\bsnm{{Magdaleni{\'c}}}, \binits{J.}}:
\byear{2024},
\batitle{{Toroidal modified Miller-Turner CME model in EUHFORIA: Validation and comparison with flux rope and spheromak}}.
\bjtitle{\aap}
\bvolume{691},
\bfpage{A146}.
\doiurl{https://doi.org/10.1051/0004-6361/202450459}.
\adsurl{2024A&A...691A.146M}.
\end{barticle}
\endbibitem

\bibitem[\protect\citeauthoryear{{Moore} et~al.}{2001}]{Moore2001}
\begin{barticle}
\bauthor{\bsnm{{Moore}}, \binits{R.L.}},
\bauthor{\bsnm{{Sterling}}, \binits{A.C.}},
\bauthor{\bsnm{{Hudson}}, \binits{H.S.}},
\bauthor{\bsnm{{Lemen}}, \binits{J.R.}}:
\byear{2001},
\batitle{{Onset of the Magnetic Explosion in Solar Flares and Coronal Mass Ejections}}.
\bjtitle{\apj}
\bvolume{552},
\bfpage{833}.
\doiurl{https://doi.org/10.1086/320559}.
\adsurl{2001ApJ...552..833M}.
\end{barticle}
\endbibitem

\bibitem[\protect\citeauthoryear{{Perri} et~al.}{2022}]{Perri2022}
\begin{barticle}
\bauthor{\bsnm{{Perri}}, \binits{B.}},
\bauthor{\bsnm{{Leitner}}, \binits{P.}},
\bauthor{\bsnm{{Brchnelova}}, \binits{M.}},
\bauthor{\bsnm{{Baratashvili}}, \binits{T.}},
\bauthor{\bsnm{{Ku{\'z}ma}}, \binits{B.}},
\bauthor{\bsnm{{Zhang}}, \binits{F.}},
\bauthor{\bsnm{{Lani}}, \binits{A.}},
\bauthor{\bsnm{{Poedts}}, \binits{S.}}:
\byear{2022},
\batitle{{COCONUT, a Novel Fast-converging MHD Model for Solar Corona Simulations: I. Benchmarking and Optimization of Polytropic Solutions}}.
\bjtitle{\apj}
\bvolume{936},
\bfpage{19}.
\doiurl{https://doi.org/10.3847/1538-4357/ac7237}.
\adsurl{2022ApJ...936...19P}.
\end{barticle}
\endbibitem

\bibitem[\protect\citeauthoryear{{Pomoell} and {Poedts}}{2018}]{Pomoell2018}
\begin{barticle}
\bauthor{\bsnm{{Pomoell}}, \binits{J.}},
\bauthor{\bsnm{{Poedts}}, \binits{S.}}:
\byear{2018},
\batitle{{EUHFORIA: European heliospheric forecasting information asset}}.
\bjtitle{Journal of Space Weather and Space Climate}
\bvolume{8},
\bfpage{A35}.
\doiurl{https://doi.org/10.1051/swsc/2018020}.
\adsurl{2018JSWSC...8A..35P}.
\end{barticle}
\endbibitem

\bibitem[\protect\citeauthoryear{{Pomoell}, {Lumme}, and {Kilpua}}{2019}]{Pomoell2019}
\begin{barticle}
\bauthor{\bsnm{{Pomoell}}, \binits{J.}},
\bauthor{\bsnm{{Lumme}}, \binits{E.}},
\bauthor{\bsnm{{Kilpua}}, \binits{E.}}:
\byear{2019},
\batitle{{Time-dependent Data-driven Modeling of Active Region Evolution Using Energy-optimized Photospheric Electric Fields}}.
\bjtitle{\solphys}
\bvolume{294},
\bfpage{41}.
\doiurl{https://doi.org/10.1007/s11207-019-1430-x}.
\adsurl{2019SoPh..294...41P}.
\end{barticle}
\endbibitem

\bibitem[\protect\citeauthoryear{{Prete} et~al.}{2024}]{Prete2024}
\begin{barticle}
\bauthor{\bsnm{{Prete}}, \binits{G.}},
\bauthor{\bsnm{{Niemela}}, \binits{A.}},
\bauthor{\bsnm{{Schmieder}}, \binits{B.}},
\bauthor{\bsnm{{Al-Haddad}}, \binits{N.}},
\bauthor{\bsnm{{Zhuang}}, \binits{B.}},
\bauthor{\bsnm{{Lepreti}}, \binits{F.}},
\bauthor{\bsnm{{Carbone}}, \binits{V.}},
\bauthor{\bsnm{{Poedts}}, \binits{S.}}:
\byear{2024},
\batitle{{EUHFORIA modelling of the Sun-Earth chain of the magnetic cloud of 28 June 2013}}.
\bjtitle{\aap}
\bvolume{683},
\bfpage{A28}.
\doiurl{https://doi.org/10.1051/0004-6361/202346906}.
\adsurl{2024A&A...683A..28P}.
\end{barticle}
\endbibitem

\bibitem[\protect\citeauthoryear{{Priest} and {Forbes}}{2002}]{Priest2002}
\begin{barticle}
\bauthor{\bsnm{{Priest}}, \binits{E.R.}},
\bauthor{\bsnm{{Forbes}}, \binits{T.G.}}:
\byear{2002},
\batitle{{The magnetic nature of solar flares}}.
\bjtitle{\aapr}
\bvolume{10},
\bfpage{313}.
\doiurl{https://doi.org/10.1007/s001590100013}.
\adsurl{2002A&ARv..10..313P}.
\end{barticle}
\endbibitem

\bibitem[\protect\citeauthoryear{{Rodriguez} et~al.}{2024}]{Rodriguez2024}
\begin{barticle}
\bauthor{\bsnm{{Rodriguez}}, \binits{L.}},
\bauthor{\bsnm{{Shukhobodskaia}}, \binits{D.}},
\bauthor{\bsnm{{Niemela}}, \binits{A.}},
\bauthor{\bsnm{{Maharana}}, \binits{A.}},
\bauthor{\bsnm{{Samara}}, \binits{E.}},
\bauthor{\bsnm{{Verbeke}}, \binits{C.}},
\bauthor{\bsnm{{Magdalenic}}, \binits{J.}},
\bauthor{\bsnm{{Vansintjan}}, \binits{R.}},
\bauthor{\bsnm{{Mierla}}, \binits{M.}},
\bauthor{\bsnm{{Scolini}}, \binits{C.}},
\bauthor{\bsnm{{Sarkar}}, \binits{R.}},
\bauthor{\bsnm{{Kilpua}}, \binits{E.}},
\bauthor{\bsnm{{Asvestari}}, \binits{E.}},
\bauthor{\bsnm{{Herbst}}, \binits{K.}},
\bauthor{\bsnm{{Lapenta}}, \binits{G.}},
\bauthor{\bsnm{{Chaduteau}}, \binits{A.D.}},
\bauthor{\bsnm{{Pomoell}}, \binits{J.}},
\bauthor{\bsnm{{Poedts}}, \binits{S.}}:
\byear{2024},
\batitle{{Validation of EUHFORIA cone and spheromak coronal mass ejection models}}.
\bjtitle{\aap}
\bvolume{689},
\bfpage{A187}.
\doiurl{https://doi.org/10.1051/0004-6361/202449530}.
\adsurl{2024A&A...689A.187R}.
\end{barticle}
\endbibitem

\bibitem[\protect\citeauthoryear{{Schmieder}}{2018}]{Schmieder2018}
\begin{barticle}
\bauthor{\bsnm{{Schmieder}}, \binits{B.}}:
\byear{2018},
\batitle{{Extreme solar storms based on solar magnetic field}}.
\bjtitle{Journal of Atmospheric and Solar-Terrestrial Physics}
\bvolume{180},
\bfpage{46}.
\doiurl{https://doi.org/10.1016/j.jastp.2017.07.018}.
\adsurl{2018JASTP.180...46S}.
\end{barticle}
\endbibitem

\bibitem[\protect\citeauthoryear{{Schmieder}, {Aulanier}, and {Vr{\v{s}}nak}}{2015}]{Schmieder2015}
\begin{barticle}
\bauthor{\bsnm{{Schmieder}}, \binits{B.}},
\bauthor{\bsnm{{Aulanier}}, \binits{G.}},
\bauthor{\bsnm{{Vr{\v{s}}nak}}, \binits{B.}}:
\byear{2015},
\batitle{{Flare-CME Models: An Observational Perspective (Invited Review)}}.
\bjtitle{\solphys}
\bvolume{290},
\bfpage{3457}.
\doiurl{https://doi.org/10.1007/s11207-015-0712-1}.
\adsurl{2015SoPh..290.3457S}.
\end{barticle}
\endbibitem

\bibitem[\protect\citeauthoryear{{Schmieder}, {Guo}, and {Poedts}}{2024}]{Schmieder2024}
\begin{barticle}
\bauthor{\bsnm{{Schmieder}}, \binits{B.}},
\bauthor{\bsnm{{Guo}}, \binits{J.}},
\bauthor{\bsnm{{Poedts}}, \binits{S.}}:
\byear{2024},
\batitle{{Recent advances in solar data-driven MHD simulations of the formation and evolution of CME flux ropes}}.
\bjtitle{Reviews of Modern Plasma Physics}
\bvolume{8},
\bfpage{27}.
\doiurl{https://doi.org/10.1007/s41614-024-00166-3}.
\adsurl{2024RvMPP...8...27S}.
\end{barticle}
\endbibitem

\bibitem[\protect\citeauthoryear{{Schmieder} et~al.}{2020}]{Schmieder2020}
\begin{barticle}
\bauthor{\bsnm{{Schmieder}}, \binits{B.}},
\bauthor{\bsnm{{Kim}}, \binits{R.-S.}},
\bauthor{\bsnm{{Grison}}, \binits{B.}},
\bauthor{\bsnm{{Bocchialini}}, \binits{K.}},
\bauthor{\bsnm{{Kwon}}, \binits{R.-Y.}},
\bauthor{\bsnm{{Poedts}}, \binits{S.}},
\bauthor{\bsnm{{D{\'e}moulin}}, \binits{P.}}:
\byear{2020},
\batitle{{Low Geo-Effectiveness of Fast Halo CMEs Related to the 12 X-Class Flares in 2002}}.
\bjtitle{Journal of Geophysical Research (Space Physics)}
\bvolume{125},
\bfpage{e27529}.
\doiurl{https://doi.org/10.1029/2019JA027529}.
\adsurl{2020JGRA..12527529S}.
\end{barticle}
\endbibitem

\bibitem[\protect\citeauthoryear{{Schrijver}}{2009}]{Schrijver2009}
\begin{barticle}
\bauthor{\bsnm{{Schrijver}}, \binits{C.J.}}:
\byear{2009},
\batitle{{Driving major solar flares and eruptions: A review}}.
\bjtitle{Advances in Space Research}
\bvolume{43},
\bfpage{739}.
\doiurl{https://doi.org/10.1016/j.asr.2008.11.004}.
\adsurl{2009AdSpR..43..739S}.
\end{barticle}
\endbibitem

\bibitem[\protect\citeauthoryear{{Schrijver} et~al.}{2012}]{Schrijver2012}
\begin{barticle}
\bauthor{\bsnm{{Schrijver}}, \binits{C.J.}},
\bauthor{\bsnm{{Beer}}, \binits{J.}},
\bauthor{\bsnm{{Baltensperger}}, \binits{U.}},
\bauthor{\bsnm{{Cliver}}, \binits{E.W.}},
\bauthor{\bsnm{{G{\"u}del}}, \binits{M.}},
\bauthor{\bsnm{{Hudson}}, \binits{H.S.}},
\bauthor{\bsnm{{McCracken}}, \binits{K.G.}},
\bauthor{\bsnm{{Osten}}, \binits{R.A.}},
\bauthor{\bsnm{{Peter}}, \binits{T.}},
\bauthor{\bsnm{{Soderblom}}, \binits{D.R.}},
\bauthor{\bsnm{{Usoskin}}, \binits{I.G.}},
\bauthor{\bsnm{{Wolff}}, \binits{E.W.}}:
\byear{2012},
\batitle{{Estimating the frequency of extremely energetic solar events, based on solar, stellar, lunar, and terrestrial records}}.
\bjtitle{Journal of Geophysical Research (Space Physics)}
\bvolume{117},
\bfpage{A08103}.
\doiurl{https://doi.org/10.1029/2012JA017706}.
\adsurl{2012JGRA..117.8103S}.
\end{barticle}
\endbibitem

\bibitem[\protect\citeauthoryear{{Scolini} et~al.}{2019}]{Scolini2019}
\begin{barticle}
\bauthor{\bsnm{{Scolini}}, \binits{C.}},
\bauthor{\bsnm{{Rodriguez}}, \binits{L.}},
\bauthor{\bsnm{{Mierla}}, \binits{M.}},
\bauthor{\bsnm{{Pomoell}}, \binits{J.}},
\bauthor{\bsnm{{Poedts}}, \binits{S.}}:
\byear{2019},
\batitle{{Observation-based modelling of magnetised coronal mass ejections with EUHFORIA}}.
\bjtitle{\aap}
\bvolume{626},
\bfpage{A122}.
\doiurl{https://doi.org/10.1051/0004-6361/201935053}.
\adsurl{2019A&A...626A.122S}.
\end{barticle}
\endbibitem

\bibitem[\protect\citeauthoryear{{Scolini} et~al.}{2020}]{scolini2020}
\begin{barticle}
\bauthor{\bsnm{{Scolini}}, \binits{C.}},
\bauthor{\bsnm{{Chan{\'e}}}, \binits{E.}},
\bauthor{\bsnm{{Pomoell}}, \binits{J.}},
\bauthor{\bsnm{{Rodriguez}}, \binits{L.}},
\bauthor{\bsnm{{Poedts}}, \binits{S.}}:
\byear{2020},
\batitle{{Improving Predictions of High-Latitude Coronal Mass Ejections Throughout the Heliosphere}}.
\bjtitle{Space Weather}
\bvolume{18},
\bfpage{e02246}.
\doiurl{https://doi.org/10.1029/2019SW002246}.
\adsurl{2020SpWea..1802246S}.
\end{barticle}
\endbibitem

\bibitem[\protect\citeauthoryear{{Shibata} et~al.}{1995}]{Shibata1995}
\begin{barticle}
\bauthor{\bsnm{{Shibata}}, \binits{K.}},
\bauthor{\bsnm{{Masuda}}, \binits{S.}},
\bauthor{\bsnm{{Shimojo}}, \binits{M.}},
\bauthor{\bsnm{{Hara}}, \binits{H.}},
\bauthor{\bsnm{{Yokoyama}}, \binits{T.}},
\bauthor{\bsnm{{Tsuneta}}, \binits{S.}},
\bauthor{\bsnm{{Kosugi}}, \binits{T.}},
\bauthor{\bsnm{{Ogawara}}, \binits{Y.}}:
\byear{1995},
\batitle{{Hot-Plasma Ejections Associated with Compact-Loop Solar Flares}}.
\bjtitle{\apjl}
\bvolume{451},
\bfpage{L83}.
\doiurl{https://doi.org/10.1086/309688}.
\adsurl{1995ApJ...451L..83S}.
\end{barticle}
\endbibitem

\bibitem[\protect\citeauthoryear{{Sturrock}}{1966}]{Sturrock1966}
\begin{barticle}
\bauthor{\bsnm{{Sturrock}}, \binits{P.A.}}:
\byear{1966},
\batitle{{Model of the High-Energy Phase of Solar Flares}}.
\bjtitle{\nat}
\bvolume{211},
\bfpage{695}.
\doiurl{https://doi.org/10.1038/211695a0}.
\adsurl{1966Natur.211..695S}.
\end{barticle}
\endbibitem

\bibitem[\protect\citeauthoryear{{Valori}, {D{\'e}moulin}, and {Pariat}}{2012}]{Valori2012}
\begin{barticle}
\bauthor{\bsnm{{Valori}}, \binits{G.}},
\bauthor{\bsnm{{D{\'e}moulin}}, \binits{P.}},
\bauthor{\bsnm{{Pariat}}, \binits{E.}}:
\byear{2012},
\batitle{{Comparing Values of the Relative Magnetic Helicity in Finite Volumes}}.
\bjtitle{\solphys}
\bvolume{278},
\bfpage{347}.
\doiurl{https://doi.org/10.1007/s11207-012-9951-6}.
\adsurl{2012SoPh..278..347V}.
\end{barticle}
\endbibitem

\bibitem[\protect\citeauthoryear{{Verbeke}, {Baratashvili}, and {Poedts}}{2022}]{verbeke2022}
\begin{barticle}
\bauthor{\bsnm{{Verbeke}}, \binits{C.}},
\bauthor{\bsnm{{Baratashvili}}, \binits{T.}},
\bauthor{\bsnm{{Poedts}}, \binits{S.}}:
\byear{2022},
\batitle{{ICARUS, a new inner heliospheric model with a flexible grid}}.
\bjtitle{\aap}
\bvolume{662},
\bfpage{A50}.
\doiurl{https://doi.org/10.1051/0004-6361/202141981}.
\adsurl{2022A&A...662A..50V}.
\end{barticle}
\endbibitem

\bibitem[\protect\citeauthoryear{{Wang} et~al.}{2025a}]{Wang2025_time}
\begin{barticle}
\bauthor{\bsnm{{Wang}}, \binits{H.P.}},
\bauthor{\bsnm{{Poedts}}, \binits{S.}},
\bauthor{\bsnm{{Lani}}, \binits{A.}},
\bauthor{\bsnm{{Brchnelova}}, \binits{M.}},
\bauthor{\bsnm{{Baratashvili}}, \binits{T.}},
\bauthor{\bsnm{{Linan}}, \binits{L.}},
\bauthor{\bsnm{{Zhang}}, \binits{F.}},
\bauthor{\bsnm{{Hou}}, \binits{D.W.}},
\bauthor{\bsnm{{Zhou}}, \binits{Y.H.}}:
\byear{2025}a,
\batitle{{Efficient magnetohydrodynamic modelling of the time-evolving corona by COCONUT}}.
\bjtitle{\aap}
\bvolume{694},
\bfpage{A234}.
\doiurl{https://doi.org/10.1051/0004-6361/202452279}.
\adsurl{2025A&A...694A.234W}.
\end{barticle}
\endbibitem

\bibitem[\protect\citeauthoryear{{Wang} et~al.}{2025b}]{Wang2025}
\begin{botherref}
\oauthor{\bsnm{{Wang}}, \binits{H.}},
\oauthor{\bsnm{{Poedts}}, \binits{S.}},
\oauthor{\bsnm{{Lani}}, \binits{A.}},
\oauthor{\bsnm{{Linan}}, \binits{L.}},
\oauthor{\bsnm{{Baratashvili}}, \binits{T.}},
\oauthor{\bsnm{{Zhang}}, \binits{F.}},
\oauthor{\bsnm{{Sorokina}}, \binits{D.}},
\oauthor{\bsnm{{Jeong}}, \binits{H.-j.}},
\oauthor{\bsnm{{Li}}, \binits{Y.}},
\oauthor{\bsnm{{Mahdi}}, \binits{N.-Z.}},
\oauthor{\bsnm{{Schmieder}}, \binits{B.}}:
2025b,
{Time-evolving coronal modelling of solar maximum around the May 2024 storm by COCONUT}.
\textit{arXiv e-prints},
arXiv:2505.11990.
\adsurl{2025arXiv250511990W}.
\end{botherref}
\endbibitem

\bibitem[\protect\citeauthoryear{{Warmuth} and {Mann}}{2020}]{Warmuth2020}
\begin{barticle}
\bauthor{\bsnm{{Warmuth}}, \binits{A.}},
\bauthor{\bsnm{{Mann}}, \binits{G.}}:
\byear{2020},
\batitle{{Thermal-nonthermal energy partition in solar flares derived from X-ray, EUV, and bolometric observations. Discussion of recent studies}}.
\bjtitle{\aap}
\bvolume{644},
\bfpage{A172}.
\doiurl{https://doi.org/10.1051/0004-6361/202039529}.
\adsurl{2020A&A...644A.172W}.
\end{barticle}
\endbibitem

\bibitem[\protect\citeauthoryear{{Yashiro} et~al.}{2005}]{Yashiro2005}
\begin{barticle}
\bauthor{\bsnm{{Yashiro}}, \binits{S.}},
\bauthor{\bsnm{{Gopalswamy}}, \binits{N.}},
\bauthor{\bsnm{{Akiyama}}, \binits{S.}},
\bauthor{\bsnm{{Michalek}}, \binits{G.}},
\bauthor{\bsnm{{Howard}}, \binits{R.A.}}:
\byear{2005},
\batitle{{Visibility of coronal mass ejections as a function of flare location and intensity}}.
\bjtitle{Journal of Geophysical Research (Space Physics)}
\bvolume{110},
\bfpage{A12S05}.
\doiurl{https://doi.org/10.1029/2005JA011151}.
\adsurl{2005JGRA..11012S05Y}.
\end{barticle}
\endbibitem

\bibitem[\protect\citeauthoryear{{Zuccarello}, {Aulanier}, and {Gilchrist}}{2015}]{Zuccarello2015}
\begin{barticle}
\bauthor{\bsnm{{Zuccarello}}, \binits{F.P.}},
\bauthor{\bsnm{{Aulanier}}, \binits{G.}},
\bauthor{\bsnm{{Gilchrist}}, \binits{S.A.}}:
\byear{2015},
\batitle{{Critical Decay Index at the Onset of Solar Eruptions}}.
\bjtitle{\apj}
\bvolume{814},
\bfpage{126}.
\doiurl{https://doi.org/10.1088/0004-637X/814/2/126}.
\adsurl{2015ApJ...814..126Z}.
\end{barticle}
\endbibitem

\bibitem[\protect\citeauthoryear{{Zuccarello} et~al.}{2018}]{Zuccarello2018}
\begin{barticle}
\bauthor{\bsnm{{Zuccarello}}, \binits{F.P.}},
\bauthor{\bsnm{{Pariat}}, \binits{E.}},
\bauthor{\bsnm{{Valori}}, \binits{G.}},
\bauthor{\bsnm{{Linan}}, \binits{L.}}:
\byear{2018},
\batitle{{Threshold of Non-potential Magnetic Helicity Ratios at the Onset of Solar Eruptions}}.
\bjtitle{\apj}
\bvolume{863},
\bfpage{41}.
\doiurl{https://doi.org/10.3847/1538-4357/aacdfc}.
\adsurl{2018ApJ...863...41Z}.
\end{barticle}
\endbibitem

\end{thebibliography}
%
%
%
%

\end{document}